\documentclass[%
 reprint,
superscriptaddress,
 amsmath,amssymb,
 aps,
prb,
floatfix,
]{revtex4-1}

\usepackage{graphicx}
\usepackage{dcolumn}
\usepackage{bm}
\usepackage{hyperref}
\usepackage[mathlines]{lineno}
\usepackage{tabularx}
\usepackage{placeins}

\begin{document}


\title{On the Origin of the Two-Dimensional Electron Gas at the CdO (100) Surface}

\author{P. C. J. Clark}
\affiliation{%
 School of Physics and Astronomy and the Photon Science Institute, The University of Manchester, Manchester M13 9PL, United Kingdom
}%


\author{A. I. Williamson}%
\affiliation{%
 School of Physics and Astronomy and the Photon Science Institute, The University of Manchester, Manchester M13 9PL, United Kingdom 
}%

\author{N. K. Lewis}%
\affiliation{%
 School of Physics and Astronomy and the Photon Science Institute, The University of Manchester, Manchester M13 9PL, United Kingdom 
}%

\author{R. Ahumada-Lazo}%
\affiliation{%
 School of Physics and Astronomy and the Photon Science Institute, The University of Manchester, Manchester M13 9PL, United Kingdom
}%


\author{M. Silly}
\affiliation{
 Synchrotron SOLEIL, BP 48, Saint-Aubin, F91192 Gif sur Yvette CEDEX, France
}%
\author{J. J. Mudd}
\affiliation{%
 Diamond Light Source, Harwell Science and Innovation Campus, Didcot OX11 ODE, United Kingdom
}%

\author{C. F. McConville}
\affiliation{%
 Department of Physics, University of Warwick, Coventry, CV4 7AL, United Kingdom
}%
\affiliation{%
 School of Science, RMIT University, Melbourne, Victoria 3001, Australia
}%

\author{W. R. Flavell}%
\affiliation{%
 School of Physics and Astronomy and the Photon Science Institute, The University of Manchester, Manchester M13 9PL, United Kingdom
}%


\date{\today}

\begin{abstract}
Synchrotron-radiation angle-resolved and core-level photoemission spectroscopy are used together to investigate the origin of the two-dimensional electron gas on the surface of single crystal CdO (100) films. A reduction in the two-dimensional electron density of the surface state is observed under the synchrotron beam during ARPES, which is shown to be accompanied by a concomitant reduction in the surface-adsorbed species (monitored through the O 1s core level signal). This shows that surface adsorbates donate electrons into the surface accumulation layer. When the surface is cleaned, the surface conduction band state empties. A surface doped with atomic H is also studied. Here, interstitial H increases the two-dimensional electron density at the surface. This demonstrates that reversible donor doping is possible. The surface band bending profiles, 2D electron densities, and effective masses are calculated from subband dispersion simulations.

\end{abstract}

\maketitle


\section{Introduction} \label{sec:Intro}

Transparent conducting oxides are of great interest for photovoltaic and optoelectronic applications as they conduct electricity whilst being optically transparent. Recently, the observation of a two-dimensional electron gas (2DEG) on the surface of metal oxides has been reported for SrTiO$_3$ \cite{Santander-Syro2011a,Meevasana2011,D'Angelo2012}, KTaO$_3$ \cite{Santander-Syro2012,Bareille2014}, anatase TiO$_2$ \cite{Rodel2015}, ZnO \cite{Piper2010,Ozawa2014}, In$_2$O$_3$ \cite{Zhang2013}, CaTiO$_3$ \cite{Muff2017}, and CdO \cite{Piper2008,King2010,Mudd2014a}. This is due to the presence of an electron accumulation layer at the surface, where the conduction band minimum dips below the charge neutrality level (CNL) at the $\Gamma$-point \cite{Piper2008}. 2DEGs allow for the study of many-body interactions in solids, and could open up new possibilities for oxide electronics. Band gap narrowing, another effect observed in oxides with quantized electron accumulation layers, offers the potential for density-controlled band-engineering of electronic devices \cite{King2010}. The high electron mobility combined with infrared transparency of CdO particularly suits the requirements needed for transparent contacts in photovoltaic devices \cite{Yu2014}. An effectively metallic state at the CdO surface could further enhance the possibilities in such structures. 

However, the exact nature of the majority donors to the electron gas on the CdO (100) surface is not well understood. Previous angle-resolved photoemission (ARPES) studies of 2DEGs on the CdO surface have been hindered in the identification of surface species by the low photon energy range of synchrotron beamlines for ARPES, where core levels could not be measured. Here we investigate the nature of donors by studying both ARPES with low photon energies, and the corresponding core levels with higher photon energies. 
We study both the as-loaded 2DEG found on the CdO surface, and one induced by exposure to atomic hydrogen. We find the intensity of the as-loaded 2DEG decreases with time when exposed to synchrotron radiation, during which surface species are removed from the surface. We show that the concentration of surface-adsorbed species is correlated with the intensity of the 2DEG, indicating that surface doping into the electron accumulation layer can be controlled by adventitious surface adsorption. We compare this 'adventitious' 2DEG with that induced by interstitial hydrogen donors \cite{Burbano2011}, implanted by hydrogen cracking the surface. Overall our work suggests that interface electronic properties of oxide heterostructures may be fine tuned by adsorption and implantation. 

\section{Experimental Details} \label{sec:Exp}

Epitaxial single crystal thin films of CdO (100) were grown on $r$-plane sapphire substrates by metal-organic vapour phase epitaxy as described previously. Structural analysis by high resolution X-ray diffraction revealed high quality (001) layers having the rock salt structure \cite{Zuniga-Perez2004}. The oxygen and cadmium precursors used were tertiary butanol and dimethylcadmium respectively. Films 400 nm thick and 5 mm$^2$ in area were studied \cite{Piper2008}. Typical bulk carrier denisities for CdO films prepared in this way range from $2.4 \times 10^{20}$ cm$^{-3}$ (as-grown) to $1-8 \times 10^{19}$ cm$^{-3}$ (ultra high vacuum (UHV) annealed) \cite{Mudd2014b,King2009c,Piper2008,Mudd2014c}. Core-level photoemission and ARPES were performed on the TEMPO beamline (50 $<$ h$\nu$ $<$ 1500 eV, equipped with a SCIENTA SES200 hemispherical analyser) at Synchrotron SOLEIL, France. Core-level spectra were recorded at room temperature in normal emission geometry, using light linearly polarized in the horizontal plane. The total instrumental resolving power $\frac{E}{\Delta E}$ is $10^4$ or better \cite{Polack2010}. ARPES and valence band measurements were taken with a photon energy of 85 eV chosen to measure around the second $\Gamma$ point \cite{Piper2008}. While this (beamline-limited) choice restricts the energy resolution attainable in ARPES, cruicially, core-level photoemission is available in the same experiment. The energy and angular resolution were 50 meV and 0.3 respectively. All core-level photoemission measurements used a photon energy selected such that the kinetic energy of the photoelectrons was approximately 75 eV, matching that of the Cd 4d quasi-core level in the VB spectra.  This was necessary to ensure that the sampling depth was similar in the core-level and valence band spectra, and both experiments were highly surface sensitive.  Core-level spectra were fitted using CasaXPS \cite{CasaXPS}. Areas extracted from the core level peaks were corrected for photon flux and photoionisation cross sections \cite{Yeh1993} before calculating ratios. Spectra were referenced to the Fermi level, measured on the sample or the Ta sample holder.

Samples were annealed in UHV by electron bombardment at 900 K following a recipe from the published literature \cite{King2009b,King2010,Mudd2014c,Mudd2014}. Atomic hydrogen cracking was performed at room temperature in UHV with a background pressure of $5 \times 10^{-8}$ mbar of molecular hydrogen passing a tungsten filament emitting electrons held at 5 A (giving a cracking efficiency estimated at  50$\%$).

\section{Computational Details} \label{sec:Comp}
The 2DEG subband dispersion was extracted by fitting Lorentzian peaks with a linear background to the energy dispersion curves (EDCs) and momentum dispersion curves (MDCs) at intervals along the dispersion curves. Peaks were fitted to both since the EDC peak is better defined (and therefore more reliable to fit) near the bottom of the band, and for MDCs definition is better at the sides \cite{Mudd2014c}. A spline fit was then made to the average the band shape. Coupled Poisson-Schr{\"o}dinger simulations\cite{King2008a} were then fitted to this spline, with the effective mass of conduction band electrons ($m_e^*$) and conduction band bending ($V_{CBB}$) as free parameters and temperature (300 K), the direct band gap (2.2 eV), the static dielectric constant (18) \cite{Finkenrath1969}, and bulk carrier density as fixed parameters. The bulk carrier density was estimated from Hall effect measurements \cite{Piper2008,Mudd2014c}. As well as $m_e^*$ and $V_{CBB}$, the simulation was also used to calculate the 2D electron density ($N_{2D}$) in the band.

\section{Results} \label{sec:Res}
\subsection{Origin of the 2DEG on as-loaded surfaces}
\begin{figure}
\includegraphics[width=\linewidth]{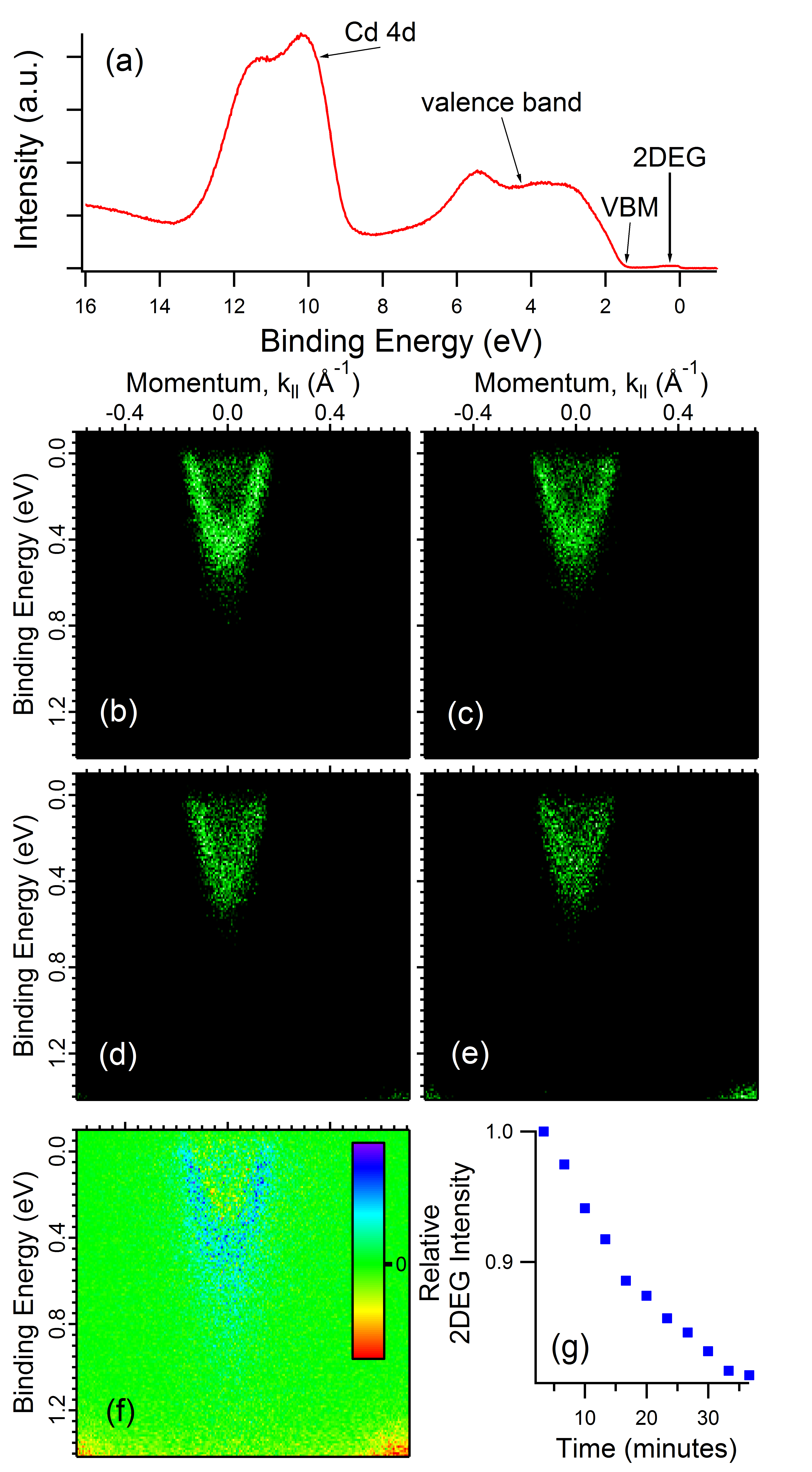} 
\caption{\label{fig:aspres} (a) valence band photoemission spectrum of the CdO surface with the main features labelled, (b)-(e) ARPES from the CdO (100) surface recorded using a photon energy of 85 eV aligned to the Fermi edge at 0 eV, showing the changes over 40 minutes' exposure to the synchrotron beam with (b) measured first and (e) last. (e) shows a difference map between the 2DEGs displayed in (b) and (e), where blue  represents positive values, red negative, and green no change;  and (g) shows the decrease in the integrated intensity of the 2DEG relative to the first scan over 36 minutes.}
\end{figure}

A valence band spectrum of the CdO surface after annealing at 900 K for 1 hour, taken at 85 eV photon energy is shown in Figure \ref{fig:aspres} (a) with the main features labelled. The feature at 10-12 eV binding energy (BE) is assigned to the Cd 4d quasi-core level \cite{King2009b}, the feature at 2-7 eV BE is assigned to the valence band, consisting of states from Cd 5s and O 2p with a small contribution from Cd 5p, and O 2s at this photon energy \cite{Mudd2014}. Close to the Fermi level (at approximately 0-0.5 eV BE) there are occupied conduction band states, with a predominately Cd 5s character and a small contribution from O 2p due to O 2p-Cd 5s hybridization \cite{Mudd2014a}. These conduction band states are occupied due to the presence of an electron accumulation layer at the surface of CdO \cite{Piper2008}. The structure of these states was studied with ARPES.

Normal emission ARPES spectra for the CdO surface after annealing at 900 K for 1 hour is presented in Figure \ref{fig:aspres} (b). The observed 2DEG has one clear subband between $\pm$ 0.18 \AA$^{-1}$ $k_{ll}$ and 0-0.6 eV BE. A less intense second subband is inside the first, between $\pm$ 0.07 \AA$^{-1}$ and 0-0.2 eV BE, but this second subband cannot be clearly distinguished. Similar subbands have been observed on the CdO surface previously \cite{Piper2008,King2010,Mudd2014c}. Decreases in the intensity and (E,$k$) width of this 2DEG were observed under the SR beam over 36 minutes, as shown by Figure \ref{fig:aspres} (b)-(e). The difference between the first and last scans is shown in Figure \ref{fig:aspres} (f). The decrease in intensity of the 2DEG with time is plotted in Figure \ref{fig:aspres} (g). The intensity was calculated by integrating the photoelectron signal from the entire 2DEG between $\pm$ 0.2 \AA$^{-1}$ and 0 to 0.8 eV BE.

\begin{figure}
\includegraphics[width=\linewidth]{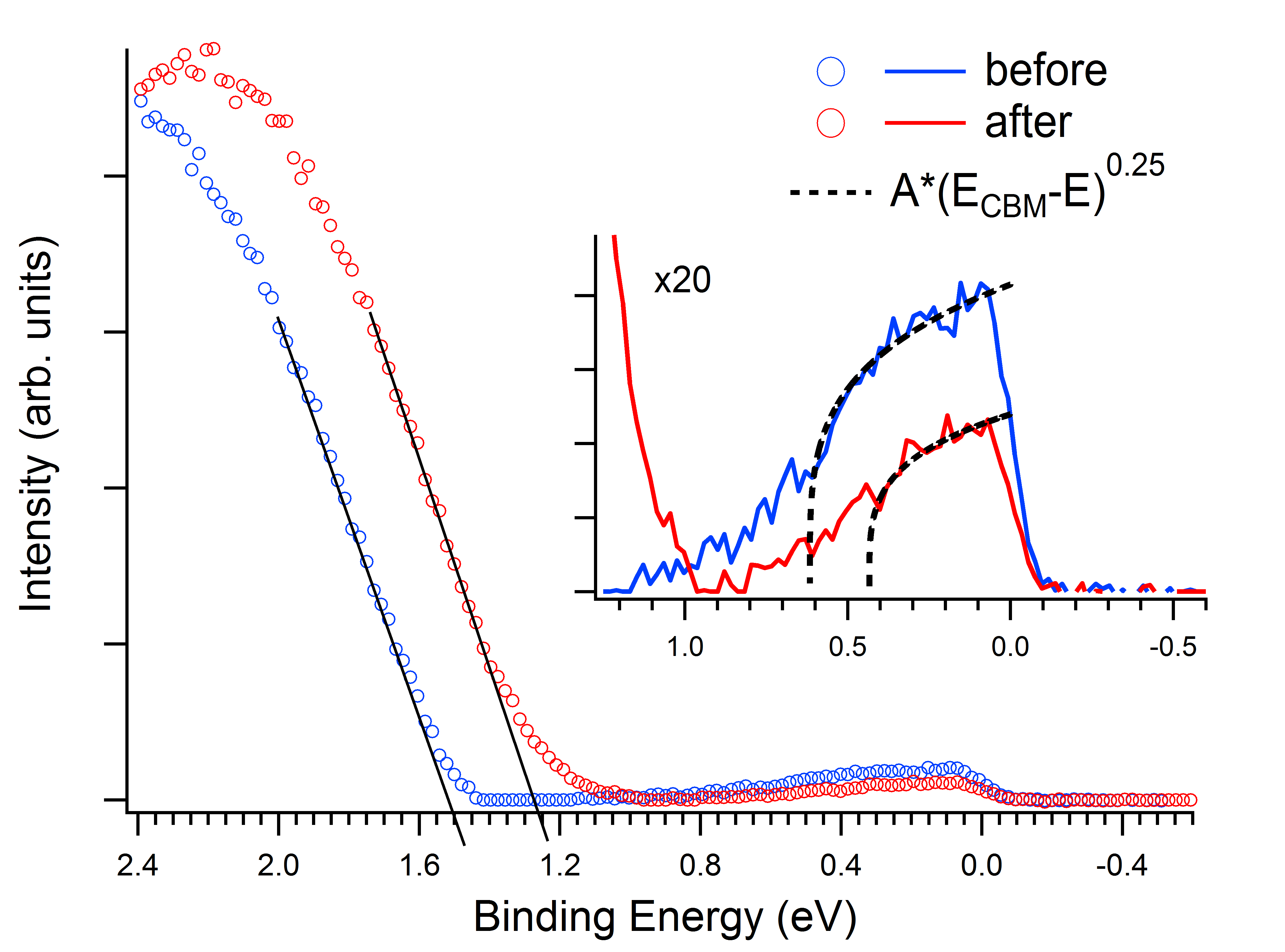} 
\caption{\label{fig:aspresinteg} Angle-integrated valence band spectra measured at 85 eV, with valence band maxima shown, for CdO (100) before and after exposure to the SR beam for 40 minutes. Inset are the spectra magnified (x20) showing the occupied states in the 2DEG. A $\frac{1}{4}$ power  function (dotted line) has been fitted to each 2DEG to estimate the BE minimum of the subband of the electron gas, where A is a constant.}
\end{figure}

The small features appearing at approximately 1.3 eV BE and $\pm$ 0.5 \AA$^{-1}$, most clearly seen in Figures \ref{fig:aspres} (e) and (f), correspond to the top of the valence band \cite{Piper2008}, implying that the valence band maximum (VBM) shifts during the experiment. This movement of the valence band to higher binding energy as the 2DEG is destroyed can be seen clearly in the angle-integrated spectra shown in Figure \ref{fig:aspresinteg}, taken before and after the ARPES data of Figure \ref{fig:aspres}. The binding energy shift in the VBM was found to be -0.24 $\pm$ 0.02 eV over the 40 minute period. As well as the decrease in intensity of the 2DEG, a shift in the BE minimum of the 2DEG can also be seen in Figure \ref{fig:aspresinteg}. To estimate this shift, a $\frac{1}{4}$ power function was fitted to the main body of the 2DEG, excluding the low intensity ``tail'' to higher binding energy, as shown by the dotted lines  in Figure \ref{fig:aspresinteg}. This function represents the dominant energy term in the nonparabolic conduction band density of states function, which has been found to best fit the conduction band dispersions of CdO \cite{King2008a,Jefferson2008,Piper2008,Mudd2014c}. The shift in the BE minimum of the 2DEG was found to be -0.19 $\pm$ 0.04 eV. Thus the shift in the VBM is the same as the shift in the BE minimum of the 2DEG, within error. Thus indicates a rigid Moss-Burstein shift in the BE minimum of the 2DEG (the conduction band minimum (CBM)) and the VBM with band filling. 

As the BE position of the VBM is not measured at the gamma point, the difference between VBM and 2DEG minimum found by this method represents the indirect bandgap. Therefore we find that the indirect band gap is constant during these measurements, and is measured to be E$_{G (indirect)}$ = 0.90 $\pm$ 0.05 eV, which agrees with the indirect band gap value of 0.89 eV calculated by Burbano \textit{et al.} via hybrid density functional theory (DFT) \citep{Burbano2011} and determined experimentally with X-ray absorption by Demchenko \textit{et al.} \citep{Demchenko2010}.

\begin{figure}
\includegraphics[width=\linewidth]{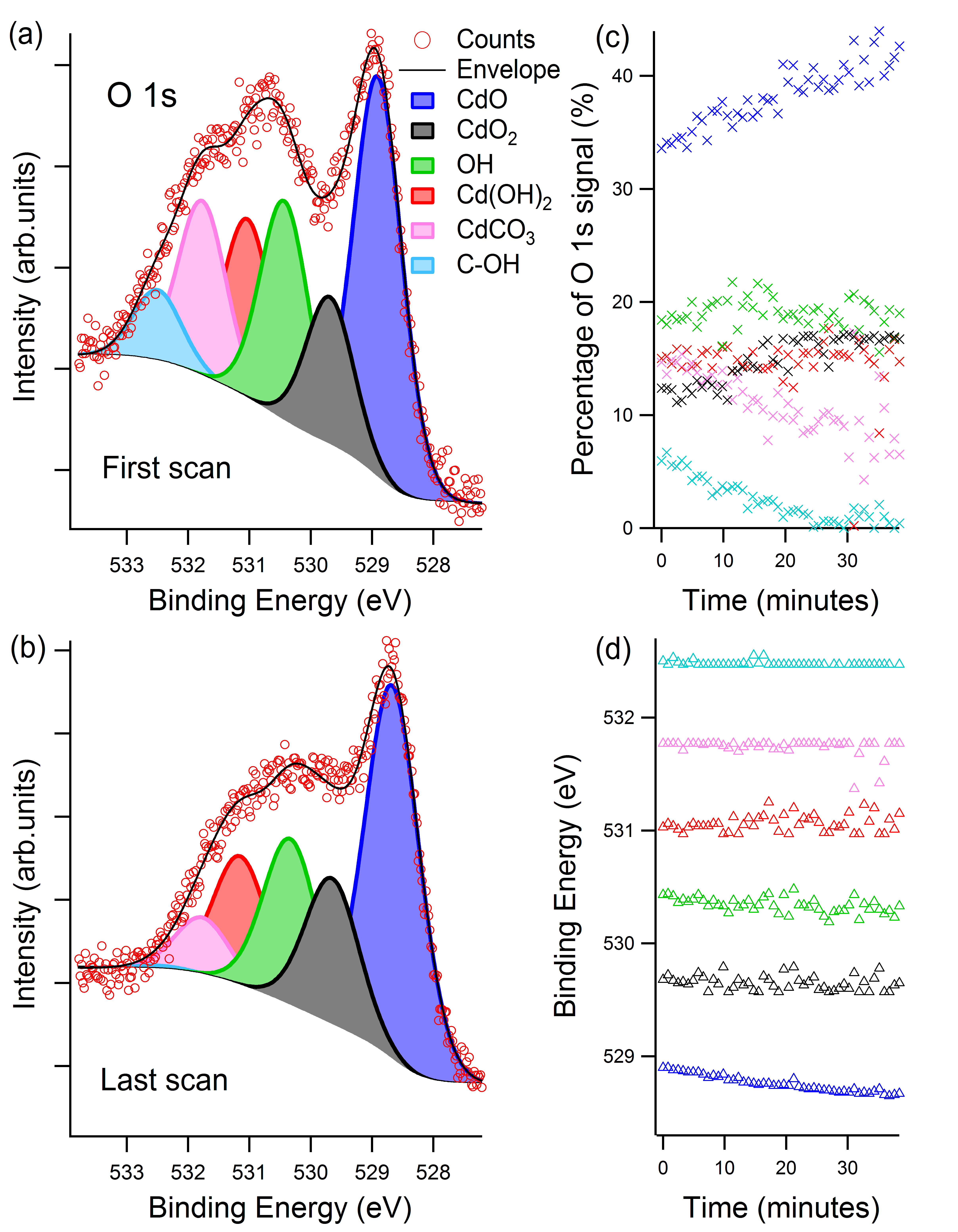} 
\caption{\label{fig:O1s} (a) and (b) Spectra of the O 1s core level showing the effect of SR exposure on the CdO (100) surface measured with a photon energy of 600 eV: (a) before exposure, (b) after 40 minutes' SR exposure, with 6 components fitted corresponding to the substrate and surface species \cite{Piper2007,Hammond1975,Maticiuc2017,Vogt2015}. Fits were performed with Gaussian-Lorentzian peaks on a Shirley background, (c) the change in composition of O 1s species with time under the SR beam, (d) the binding energy positions of the different components as a function of time.}
\end{figure}

To investigate why these changes in the 2DEG occur, the species present at the surface were investigated using core-level photoemission recorded over similar time periods and using similar flux to the ARPES measurements. The O 1s spectrum was measured repeatedly over 40 minutes. The photon fluxes 
were $2.4 \times 10^{12}$ and $2.0 \times 10^{12}$ photons/second at 600 eV (O 1s measurement) and 85 eV (ARPES measurement) respectively, i.e. the flux is reasonably similar between the two measurements. 
The spectra at the start and end of the experiment are shown in Figures \ref{fig:O1s} (a) and (b). Peaks were fitted with a binding energy position restriction of $\pm$ 0.2 eV; assignments are displayed in table 1. The main peak, at lowest binding energy (528.9 eV), is assigned to CdO \cite{Piper2007,Hammond1975}. Next to this, at 529.6 eV BE, is a peak previously attributed to CdO$_2$ \cite{Piper2007,Hammond1975}. The components at 530.2 and 530.9 eV BE are associated with strongly and loosely bound hydroxides respectively \cite{Hammond1975,Maticiuc2017,Vogt2015}. CdCO$_3$ is located at 531.6 eV BE and finally the peak at 532.5 eV BE is associated with residual alcohols present from the growth precursors \cite{Zuniga-Perez2004,Piper2007,Hammond1975}. Thus, significant amounts of surface adsorbates remain at the surface after the usually adopted cleaning procedure (annealing in UHV for 1 hour at 900 K). 

Figure \ref{fig:O1s} (c) shows that the intensity of the peaks from the oxides, CdO and CdO$_2$ \cite{Piper2007,Hammond1975}, increase relative to the other peaks with time during the experiment. The peak originating from C-OH is almost completely removed after 40 minutes \cite{Piper2007}. CdCO$_3$ is also removed under the beam. To check that the different photon energies used to measure the ARPES and O 1s were not affecting the desorption, we also examined the O 1s spectra measured before and after ARPES measurements (supplementary information). The change in surface species is consistent with those in Figure \ref{fig:O1s}. The removal of these components from the surface of CdO occurs in parallel with the reduction in intensity of the 2DEG. This suggests that the adsorbates donate electrons to the 2DEG on the CdO surface. 

Figure \ref{fig:O1s} (d) shows the change in the binding energy positions of the different components with time. The most obvious change is in the CdO peak position, which moves to lower binding energy by 0.23 $\pm$ 0.02 eV, consistent with the shifts observed in the VBM and CBM, and reflecting the Moss-Burstein shift \cite{King2009a}. 

\begin{table}
\caption{\label{tab:O1s} XPS assignments for the O 1s region.}
\begin{ruledtabular}
\begin{tabular}{lll}
Component & BE (eV) & $\Delta$ BE from CdO (eV) \\
CdO & 528.9 \cite{Piper2007,Hammond1975} & -\\
CdO$_2$ & 529.6 \cite{Piper2007,Hammond1975} & +0.7\\
OH & 530.2 \cite{Maticiuc2017} & +1.3\\
Cd(OH)$_2$ & 530.9 \cite{Hammond1975,Vogt2015} & +2\\
CdCO$_3$ & 531.6 \cite{Piper2007,Hammond1975} & +2.5\\
C-OH & 532.5 \cite{Piper2007} & +3.6\\
\end{tabular}
\end{ruledtabular}
\end{table}

The 2DEG intensity and shape was also found to vary with position over the CdO surface. This we believe is due to variation in the concentration of surface adsorbates with position due to inhomogeneous annealing temperatures across the surface. To check this observation is not due to varying topography of the surface SEM and EDX were performed (see supplementary information). EDX revealed a homogeneous composition consistent with CdO (within the limits of the technique), while SEM showed the topography to be reasonably uniform across the surface.

\begin{figure}
\includegraphics[width=\linewidth]{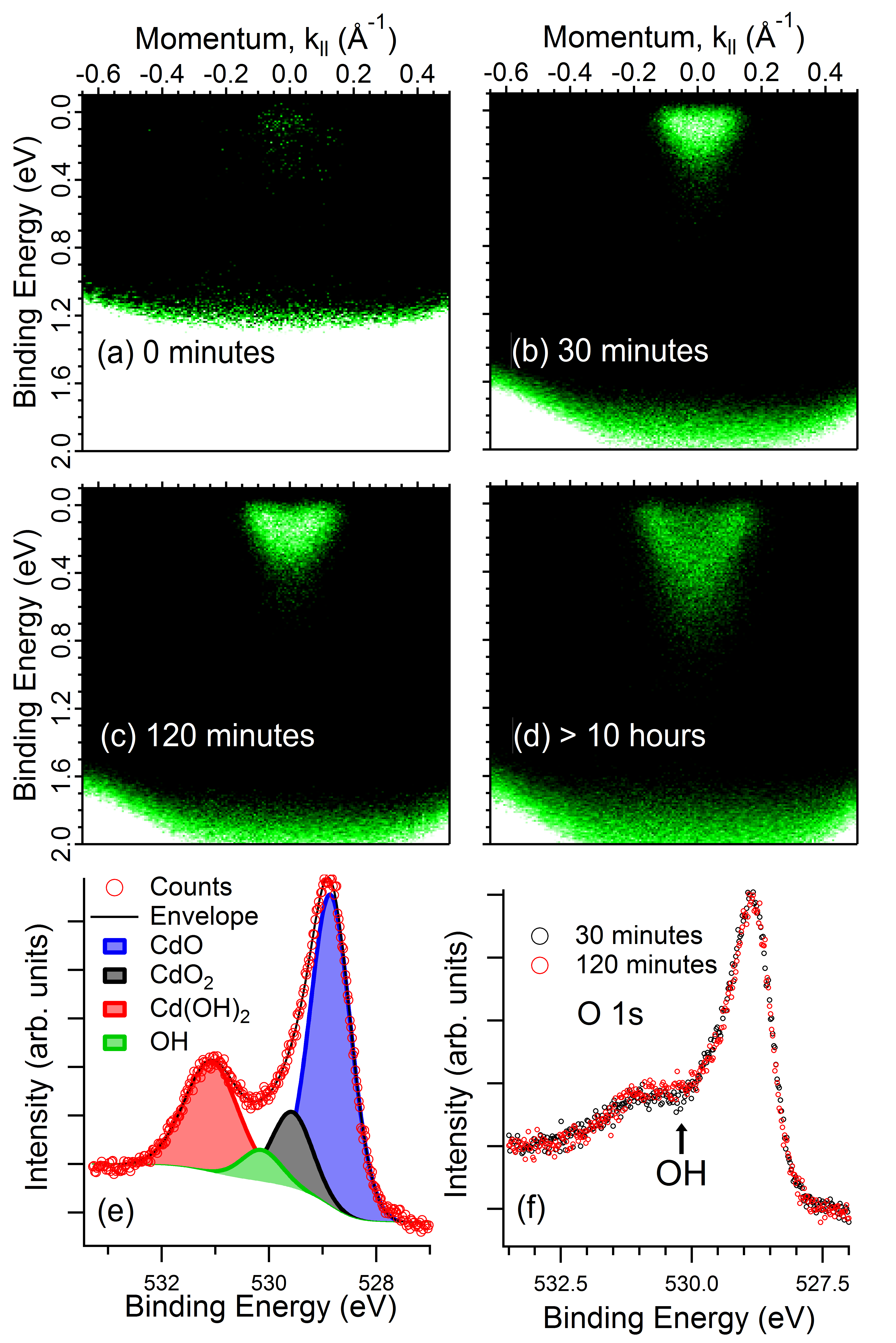} 
\caption{\label{fig:Hcrack} (a) – (d) ARPES measurements from the CdO (100) surface, recorded using a photon energy of 85 eV aligned to the Fermi edge at 0.0 eV: (a) clean and (b)-(d) after increasing exposure to atomic H, (e) and (f):  XPS spectra of the O 1s core level for (e) the clean surface and (f) 30 and 120 minutes of atomic H exposure.}
\end{figure}

To confirm the hypothesis that these surface species are responsible for the 2DEG observed at the CdO surface we annealed the surface in UHV for a further 2.5 h at 900 K (in addition to 1 h at 900 K initially, i.e. 3.5 h total) until complete removal of the C species was observed in the C 1s region (supplementary information). ARPES after this treatment is shown in Figure \ref{fig:Hcrack} (a), which indicates that the states at the Fermi level due to the 2DEG are almost completely removed. A weak intensity state is still observed, which we attribute to electron donation from hydroxides still present on the surface (as some residual Cd(OH)$_2$ and OH species are still observed in the O 1s spectrum, Figure \ref{fig:Hcrack}(e)). The idea that doping to/from the accumulation layer can be achieved via surface adsorbates is consistent with studies on In$_2$O$_3$ \cite{Berthold2016}. Binding energy shifts in core levels of oxides such as TiO$_2$ due to surface adsorbates changing the band bending are also commonly observed \cite{Thomas2012}. Our work suggests that band filling in the 2DEG observed in the as-loaded CdO (100) surface is strongly influenced by donation from surface-adsorbed species.

\subsection{Hydrogen Cracking - Creating a 2DEG by donor implantation}

Starting with the cleanest possible surface, we then investigated the creation of a 2DEG through atomic hydrogen cracking above the surface. ARPES was measured after hydrogen cracking treatment for 30 minutes, 120 minutes, and $>$ 10 hours as described above. The progressive formation of a 2DEG with prolonged exposure to atomic hydrogen can be seen in Figure \ref{fig:Hcrack} (b)-(d). After 30 minutes, the width of the state in k-space at the Fermi energy is approximately 0.2 \AA$^{-1}$ and the energy bandwidth is 0.27 eV. After a further 90 minutes, these parameters are 0.28 \AA$^{-1}$ and 0.31 eV.  The characteristic nonparabolic ``V'' shape of the state emerges \cite{Piper2008,King2010}. After $>$ 10 h the 2DEG begins to split into two subbands and to resemble the shape seen in Figure \ref{fig:aspres} (b). A Moss-Burstein shift \cite{King2009a} can also be seen in the ARPES, particularly before and after the first 30 minutes of H exposure as the energy minimum of the 2DEG is shifted to higher binding energy relative to the Fermi level. The VBM and CBM shift by the same amount (0.35 $\pm$ 0.05 eV).

The comparison between the O 1s signal before (Figure \ref{fig:Hcrack} (e)) and after H exposure (Figure \ref{fig:Hcrack} (f)) shows that some surface species desorb during the exposure to atomic hydrogen.  However, this appears to occur at the start of the cracking process; by comparing the O 1s spectra obtained at 30 and 120 minute cracking times, shown in Figure \ref{fig:Hcrack} (f), we find only a small change in the oxygen species over the intervening period. This change is a small increase in the OH component intensity (4.7$\%$ to 7.8$\%$ of the total signal for 30 minutes and 120 minutes cracking times respectively). Hydrogen cracking surfaces by this method is known to create interstitial hydrogen with implantation depths greater than the sampling depths of our experiment \cite{Ozawa2011,King2009a,McCluskey2009,Brillson2007,Lord2017,Nickel2006}. As the interstitial H is positioned close to O in the CdO lattice \cite{Burbano2011}, this OH component increase may be attributed to increased amounts of interstitial H in the lattice. A similar assignment has been made from the O 2s spectra of hydrogen-cracked ZnO \cite{Ozawa2011}, and is discussed in more detail later. 

These observations suggest that during cracking, the doping of the 2DEG is controlled primarily by the implantation of interstitial H \cite{Burbano2011,King2009a}, and not by a change in the surface species. Some damage to this 2DEG was also observed under the SR beam through a reduction in intensity over time for the $>$ 10 h H-cracked surface. Again the O 1s spectrum was used as an indication of any change in surface species, and a reduction in the OH peak intensity was found after 1 hour under the SR beam (supplementary information), suggesting, as for the as-loaded surface, the removal of donors. 

\subsection{Subband Dispersion Simulations}
The dispersion was simulated for the annealed surface at three intervals during the 40 minutes of synchrotron beam exposure. The best agreement between the observed subbands in the ARPES and the simulated dispersions was achieved using a bulk carrier density of $7 \times 10^{19}$ cm$^{-3}$. A value of $m_e^*$ = 0.18$m_e$ for the first scan (starting at 0 minutes exposure) was found, and this value was used as the conduction band electron effective for subsequent simulations. The calculated values of $V_{CBB}$ (the conduction band band bending) and $N_{2D}$ for each exposure time are shown in Table \ref{tab:sim1}. The valence band bending ($V_{VBB}$) values are also given for times 0 and 40 minutes in Table \ref{tab:sim1}, calculated with
\begin{equation}
V_{VBB} = \epsilon - E_{G (indirect)} - E_{CB-F} \label{eq:VVBB}
\end{equation}

where $\epsilon$ is the VB edge extracted from Figure \ref{fig:aspresinteg}, $E_{G (indirect)}$ is the indirect band gap calculated earlier (0.9 eV), and the offset between the bulk CBM and $E_F$, $E_{CB-F}$ = 0.1 eV \cite{Mudd2014c}. 

\begin{table*}
\caption{\label{tab:sim1} Numerical results obtained from fitting subband dispersions at 3 times during the synchrotron radiation exposure using the coupled Poisson-Schr{\"o}dinger approach. $V_{VBB}$ is calculated from experimental values, and $V_{CBB}$ and $N_{2D}$ are calculated from the subband dispersion simulations. $^a$The valence band edge was not measured at this time so $V_{VBB}$ could not be calculated.}

\begin{ruledtabular}
\begin{tabular}{llll}
Time under SR beam (minutes) & 0 & 20 & 40 \\
$V_{VBB}$ (eV) & $0.50 \pm 0.05$ & -$^a$	& $0.26 \pm 0.05$ \\
$V_{CBB}$ (eV) & $0.66 \pm 0.05$ & $0.53 \pm 0.05$	& $0.42 \pm 0.05$ \\
$N_{2D}$ $(\times 10^{13}$ cm$^{-2}$) & $4.2 \pm 0.1$ & $3.3 \pm 0.1$	& $2.5 \pm 0.1$ \\
\end{tabular}
\end{ruledtabular}
\end{table*}

From Table \ref{tab:sim1} we can see that $V_{CBB}$ and $N_{2D}$ decrease under the synchrotron beam, as expected by the change in shape and decrease in intensity of the 2DEG seen in Figure \ref{fig:aspres}. 
Band bending profiles for the CB and VB for 0 minutes and 40 minutes’ synchrotron radiation exposure are plotted in Figure 5. The electron densities (Table \ref{tab:sim1}) and calculated band bending profiles are similar to those previously found for CdO, and a small amount of surface band gap narrowing (-0.16 $\pm$ 0.5 eV for the 0 minutes exposure data) is observed compared to the bulk band gap \cite{Piper2008,King2010,Mudd2014c}. 

\begin{figure}
\includegraphics[width=\linewidth]{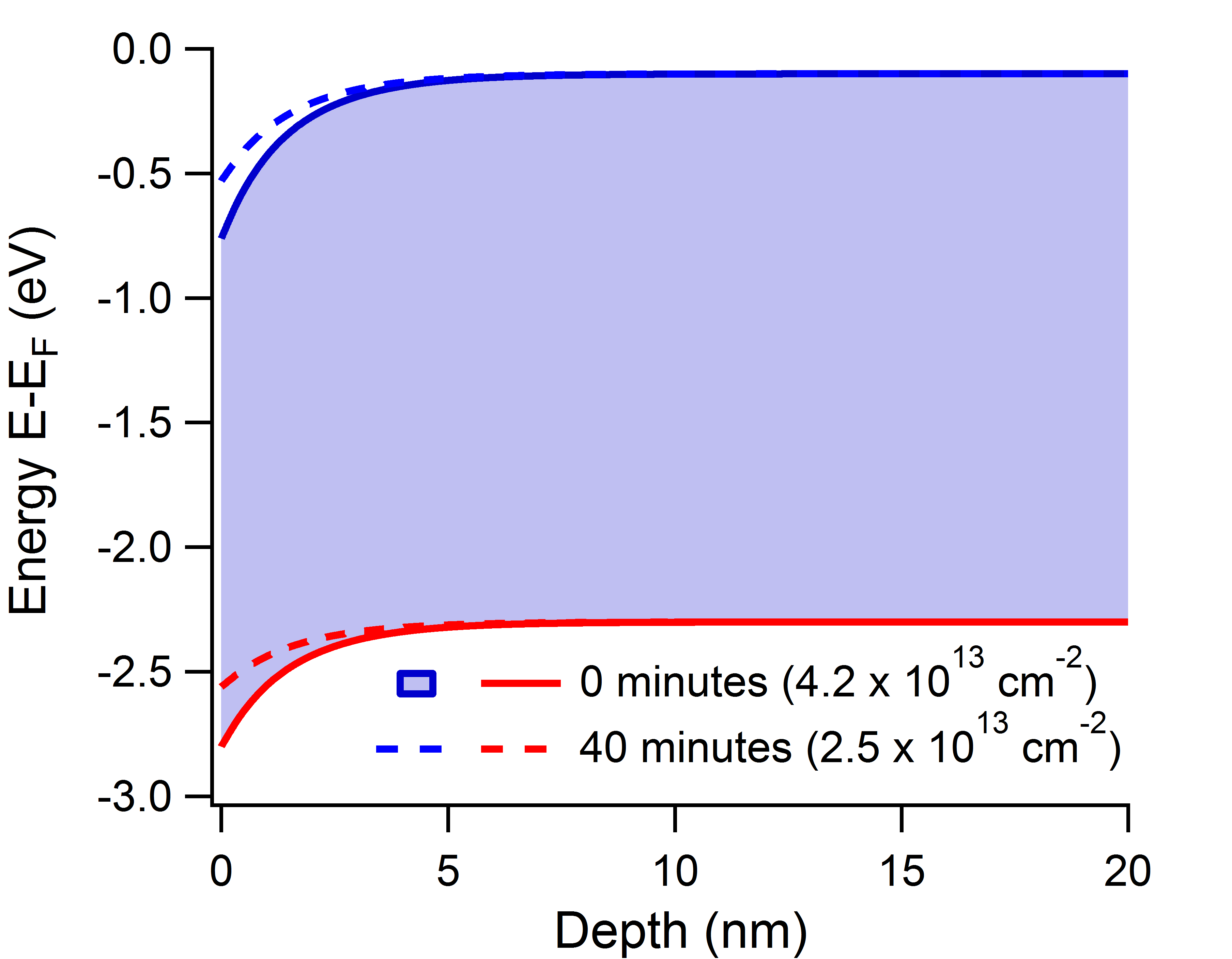} 
\caption{\label{fig:bbprofiles} Band bending profiles for the CB and VB near the CdO surface (0 nm) at two times during 85 eV synchrotron radiation exposure. The 2D electron densities from simulation of the measured subband dispersions are shown in brackets. The valence band bendings are calculated from experimental values, and the conduction band bendings are calculated from the subband dispersion simulations.}
\end{figure}

For the H-cracked surface, the ARPES data subband dispersions could not be fitted since the subband structure is not distinctly resolved for those dopant concentrations (Figure \ref{fig:Hcrack}). Instead the 2D electron density was estimated using
\begin{equation}
N_{2D} = \frac{k_F^2}{2\pi} \label{eq:N2D}
\end{equation}
where $k_F$ is the width of the observed 2DEG. The width was taken from the centre of the peak in the MDC of the ARPES data near the Fermi level (see supplementary information). The width and calculated 2D electron densities from equation \ref{eq:N2D} for each hydrogen cracking time are shown in Table \ref{tab:sim2}. 

\begin{table*}
\caption{\label{tab:sim2} Two dimensional electron densities calculated for the H doped CdO surface at various times during hydrogen cracking.}
\begin{ruledtabular}
\begin{tabular}{llll}
Cracking time (hours) & 0.5 & 2 & $>$10 \\
$k_F$ (\AA $^{-1}$) & $0.05 \pm 0.01$ & $0.08 \pm 0.01$	& $0.13 \pm 0.01$ \\
$N_{2D}$ $(\times 10^{13}$ cm$^{-2}$) & $0.4 \pm 0.1$ & $1.0 \pm 0.1$	& $2.7 \pm 0.1$ \\
\end{tabular}
\end{ruledtabular}
\end{table*}

\section{Discussion} \label{sec:Discussion}
The combination of angle-resolved and core-level photoemission data allows for a fuller understanding of the changes observed with time both under synchrotron radiation exposure and with hydrogen cracking. This has not previously been studied for CdO, and has allowed us to determine that the band filling of the accumulation layer on the as-prepared surface is strongly influenced by electron donation from surface adsorbates. Removal of surface adsorbates correlates well with the reduction of the 2D electron densities in the surface states. When the surface is annealed to remove as many adsorbates as possible, the surface state is almost completely removed. 

We also used this combination of photoemission techniques to study the creation of a 2DEG at the CdO surface by hydrogen cracking. 
The interstitial H implanted into the CdO surface during hydrogen cracking is positioned close to O ions in the lattice, and can therefore be likened to OH \cite{Burbano2011}. Similar experiments on ZnO show that H bonds to O atoms during hydrogen cracking at room temperature \cite{Ozawa2011,Wang2005a}. Therefore doping the CdO surface with H is not dissimilar to doping with surface-adsorbed hydroxide species. Indeed Ozawa \textit{et al.} produced similar subbands in the accumulation layer on the ZnO ($10\bar{1}0$) surface by hydrogen cracking and by the adsorption of methanol and water \cite{Ozawa2010}.

King \textit{et al.} previously studied the incorporation of hydrogen defects in CdO and observed an increased Burstein-Moss shift in the optical absorption edge and increased bulk carrier concentrations after hydrogen diffusion. They attributed these changes to the donor nature of hydrogen in n-type CdO \cite{King2009a}. Our results here corroborate this, as clearly interstitial H increases the intensity of the 2DEG we observe with ARPES, and lowers the BE minimum of the 2DEG (Figure \ref{fig:Hcrack} (a-d)).

The position of the CNL relative to the CBM is responsible for the donor behaviour of the interstitial H and surface species \cite{King2011}. When the CBM is below the CNL energy, virtual in-gap states become predominantly donor-like, and when above it, they become acceptor-like. Through DFT calculations, the CNL has previously been found to lie above the CBM at the $\Gamma$-point in CdO, since the CBM is much lower here compared to the rest of the Brillouin zone \cite{Piper2008}. Due to this, all defects are likely to act as donors in CdO \cite{Burbano2011}, and hence to donate electrons to the pocket of states below the Fermi energy at the $\Gamma$-point.

Subband dispersion simulations were used to find the 2D electron density, CB electron effective mass and band bending magnitudes for the as-prepared surfaces. The electron effective mass for the CB found by the simulations (0.18$m_e$) is similar to, although slightly lower than those calculated by Burbano \textit{et al.} with DFT (0.21 $m_e$) \cite{Burbano2011}. Lower effective masses have been calculated for the subband states on CdO previously \cite{Mudd2014c}, and significant many-body effects at the surface were suggested as the cause of the reduced effective masses, although the mechanism is not understood.

\begin{figure}
\includegraphics[width=\linewidth]{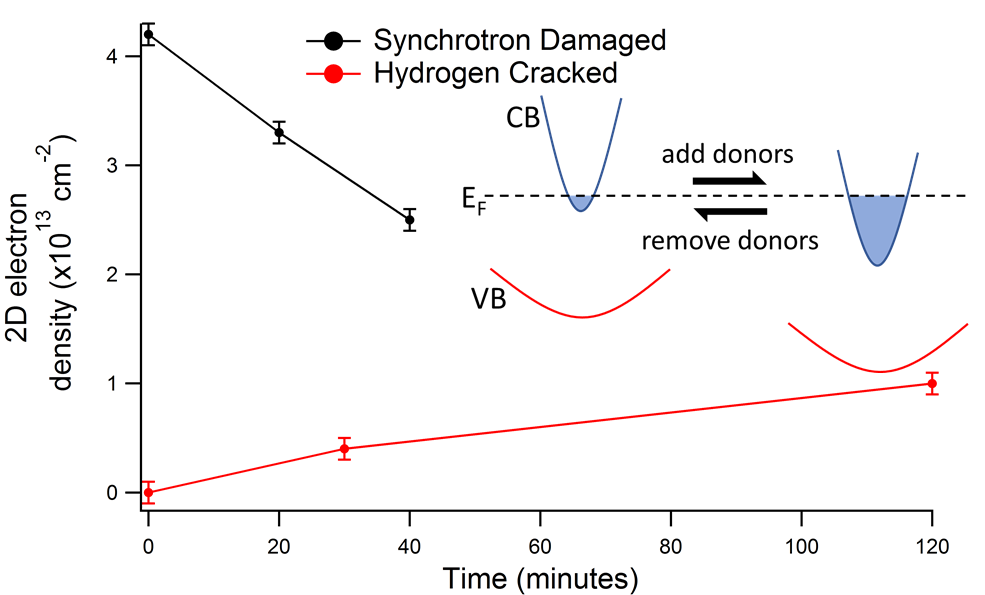} 
\caption{\label{fig:2Dedensitiesvstime3} Two dimensional electron densities for the as-loaded surface damaged by synchrotron radiation (black) and the hydrogen cracked surface (red) with time under synchrotron radiation exposure and hydrogen cracking respectively. The hydrogen cracked data starts with the clean, undoped surface. A schematic diagram of the bands is inset to demonstrate the reversibility of adding and removing donors and how this effects the band positions.}
\end{figure}

Tables \ref{tab:sim1} and \ref{tab:sim2} show how the 2D electron density at the surface of CdO changes by adding or removing dopants. With synchrotron radiation exposure, the electron density decreases as dopants are effectively lost from the surface \textit{via} adsorbate removal. With hydrogen cracking, interstitial hydrogen increases the electron density. These effects are plotted against time in Figure \ref{fig:2Dedensitiesvstime3}. In both cases the donors are hydroxide-like, and thus the doping is reversible. The inset diagram in Figure \ref{fig:2Dedensitiesvstime3} highlights the reversibility of doping in and out of the 2DEG and the shift of the conduction and valence bands at the surface relative to the Fermi level. 

\FloatBarrier
\section{Conclusion} \label{sec:Conc}

The origins of the 2DEG occurring on both as-loaded, and hydrogen doped CdO (100) surfaces have been investigated. For the as-loaded surface, we find that species present on the surface such as alcohols and hydroxides have a strong influence on the population of the 2DEG, donating electrons into the CdO surface. The use of core-level photoemission to investigate the O 1s composition changing with time was crucial to this understanding. For the 2DEG produced through exposure to atomic hydrogen, we suggest interstitial hydrogen is the donor. In both cases the 2DEG is damaged under the SR beam, seen as a reduction in intensity and change in shape. This damage is due to the removal of the donor species. 

Simulations of the surface subband dispersions were performed to calculate the band bending profiles, CB electron effective mass and 2D electron carrier densities. These results are similar to previously determined values. We have demonstrated that the 2D electron density at the surface of CdO can be controlled by the addition/removal of donors, with both surface adsorbates and interstitial hydrogen. Understanding how to control the 2D electron density and other properties of surface/interface states is crucial to the development of oxide electronics. Our work provides alternative routes for tailoring the doping level and electronic properties in interface 2DEGs in oxide heterostructures.

\begin{acknowledgments}
The research leading to these results has been supported by the project CALIPSOplus under the Grant Agreement 730872 from the EU Framework Programme for Research and Innovation HORIZON 2020. The data associated with this article are openly available from Mendeley Data, DOI: 10.17632/n59c4s84k8.1.
\end{acknowledgments}

\bibliography{CdOpaper}

\begin{thebibliography}{43}%
\makeatletter
\providecommand \@ifxundefined [1]{%
 \@ifx{#1\undefined}
}%
\providecommand \@ifnum [1]{%
 \ifnum #1\expandafter \@firstoftwo
 \else \expandafter \@secondoftwo
 \fi
}%
\providecommand \@ifx [1]{%
 \ifx #1\expandafter \@firstoftwo
 \else \expandafter \@secondoftwo
 \fi
}%
\providecommand \natexlab [1]{#1}%
\providecommand \enquote  [1]{``#1''}%
\providecommand \bibnamefont  [1]{#1}%
\providecommand \bibfnamefont [1]{#1}%
\providecommand \citenamefont [1]{#1}%
\providecommand \href@noop [0]{\@secondoftwo}%
\providecommand \href [0]{\begingroup \@sanitize@url \@href}%
\providecommand \@href[1]{\@@startlink{#1}\@@href}%
\providecommand \@@href[1]{\endgroup#1\@@endlink}%
\providecommand \@sanitize@url [0]{\catcode `\\12\catcode `\$12\catcode
  `\&12\catcode `\#12\catcode `\^12\catcode `\_12\catcode `\%12\relax}%
\providecommand \@@startlink[1]{}%
\providecommand \@@endlink[0]{}%
\providecommand \url  [0]{\begingroup\@sanitize@url \@url }%
\providecommand \@url [1]{\endgroup\@href {#1}{\urlprefix }}%
\providecommand \urlprefix  [0]{URL }%
\providecommand \Eprint [0]{\href }%
\providecommand \doibase [0]{http://dx.doi.org/}%
\providecommand \selectlanguage [0]{\@gobble}%
\providecommand \bibinfo  [0]{\@secondoftwo}%
\providecommand \bibfield  [0]{\@secondoftwo}%
\providecommand \translation [1]{[#1]}%
\providecommand \BibitemOpen [0]{}%
\providecommand \bibitemStop [0]{}%
\providecommand \bibitemNoStop [0]{.\EOS\space}%
\providecommand \EOS [0]{\spacefactor3000\relax}%
\providecommand \BibitemShut  [1]{\csname bibitem#1\endcsname}%
\let\auto@bib@innerbib\@empty
\bibitem [{\citenamefont {Santander-Syro}\ \emph {et~al.}(2011)\citenamefont
  {Santander-Syro}, \citenamefont {Copie}, \citenamefont {Kondo}, \citenamefont
  {Fortuna}, \citenamefont {Pailh{\`{e}}s}, \citenamefont {Weht}, \citenamefont
  {Qiu}, \citenamefont {Bertran}, \citenamefont {Nicolaou}, \citenamefont
  {Taleb-Ibrahimi}, \citenamefont {{Le F{\`{e}}vre}}, \citenamefont {Herranz},
  \citenamefont {Bibes}, \citenamefont {Reyren}, \citenamefont {Apertet},
  \citenamefont {Lecoeur}, \citenamefont {Barth{\'{e}}l{\'{e}}my},\ and\
  \citenamefont {Rozenberg}}]{Santander-Syro2011a}%
  \BibitemOpen
  \bibfield  {author} {\bibinfo {author} {\bibfnamefont {A.~F.}\ \bibnamefont
  {Santander-Syro}}, \bibinfo {author} {\bibfnamefont {O.}~\bibnamefont
  {Copie}}, \bibinfo {author} {\bibfnamefont {T.}~\bibnamefont {Kondo}},
  \bibinfo {author} {\bibfnamefont {F.}~\bibnamefont {Fortuna}}, \bibinfo
  {author} {\bibfnamefont {S.}~\bibnamefont {Pailh{\`{e}}s}}, \bibinfo {author}
  {\bibfnamefont {R.}~\bibnamefont {Weht}}, \bibinfo {author} {\bibfnamefont
  {X.~G.}\ \bibnamefont {Qiu}}, \bibinfo {author} {\bibfnamefont
  {F.}~\bibnamefont {Bertran}}, \bibinfo {author} {\bibfnamefont
  {A.}~\bibnamefont {Nicolaou}}, \bibinfo {author} {\bibfnamefont
  {A.}~\bibnamefont {Taleb-Ibrahimi}}, \bibinfo {author} {\bibfnamefont
  {P.}~\bibnamefont {{Le F{\`{e}}vre}}}, \bibinfo {author} {\bibfnamefont
  {G.}~\bibnamefont {Herranz}}, \bibinfo {author} {\bibfnamefont
  {M.}~\bibnamefont {Bibes}}, \bibinfo {author} {\bibfnamefont
  {N.}~\bibnamefont {Reyren}}, \bibinfo {author} {\bibfnamefont
  {Y.}~\bibnamefont {Apertet}}, \bibinfo {author} {\bibfnamefont
  {P.}~\bibnamefont {Lecoeur}}, \bibinfo {author} {\bibfnamefont
  {A.}~\bibnamefont {Barth{\'{e}}l{\'{e}}my}}, \ and\ \bibinfo {author}
  {\bibfnamefont {M.~J.}\ \bibnamefont {Rozenberg}},\ }\href {\doibase
  10.1038/nature09720} {\bibfield  {journal} {\bibinfo  {journal} {Nature}\
  }\textbf {\bibinfo {volume} {469}},\ \bibinfo {pages} {189} (\bibinfo {year}
  {2011})}\BibitemShut {NoStop}%
\bibitem [{\citenamefont {Meevasana}\ \emph {et~al.}(2011)\citenamefont
  {Meevasana}, \citenamefont {King}, \citenamefont {He}, \citenamefont {Mo},
  \citenamefont {Hashimoto}, \citenamefont {Tamai}, \citenamefont
  {Songsiriritthigul}, \citenamefont {Baumberger},\ and\ \citenamefont
  {Shen}}]{Meevasana2011}%
  \BibitemOpen
  \bibfield  {author} {\bibinfo {author} {\bibfnamefont {W.}~\bibnamefont
  {Meevasana}}, \bibinfo {author} {\bibfnamefont {P.~D.~C.}\ \bibnamefont
  {King}}, \bibinfo {author} {\bibfnamefont {R.~H.}\ \bibnamefont {He}},
  \bibinfo {author} {\bibfnamefont {S.-k.}\ \bibnamefont {Mo}}, \bibinfo
  {author} {\bibfnamefont {M.}~\bibnamefont {Hashimoto}}, \bibinfo {author}
  {\bibfnamefont {A.}~\bibnamefont {Tamai}}, \bibinfo {author} {\bibfnamefont
  {P.}~\bibnamefont {Songsiriritthigul}}, \bibinfo {author} {\bibfnamefont
  {F.}~\bibnamefont {Baumberger}}, \ and\ \bibinfo {author} {\bibfnamefont
  {Z.-x.}\ \bibnamefont {Shen}},\ }\href {\doibase 10.1038/NMAT2943} {\bibfield
   {journal} {\bibinfo  {journal} {Nature Materials}\ }\textbf {\bibinfo
  {volume} {10}},\ \bibinfo {pages} {114} (\bibinfo {year} {2011})}\BibitemShut
  {NoStop}%
\bibitem [{\citenamefont {D'Angelo}\ \emph {et~al.}(2012)\citenamefont
  {D'Angelo}, \citenamefont {Yukawa}, \citenamefont {Ozawa}, \citenamefont
  {Yamamoto}, \citenamefont {Hirahara}, \citenamefont {Hasegawa}, \citenamefont
  {Silly}, \citenamefont {Sirotti},\ and\ \citenamefont
  {Matsuda}}]{D'Angelo2012}%
  \BibitemOpen
  \bibfield  {author} {\bibinfo {author} {\bibfnamefont {M.}~\bibnamefont
  {D'Angelo}}, \bibinfo {author} {\bibfnamefont {R.}~\bibnamefont {Yukawa}},
  \bibinfo {author} {\bibfnamefont {K.}~\bibnamefont {Ozawa}}, \bibinfo
  {author} {\bibfnamefont {S.}~\bibnamefont {Yamamoto}}, \bibinfo {author}
  {\bibfnamefont {T.}~\bibnamefont {Hirahara}}, \bibinfo {author}
  {\bibfnamefont {S.}~\bibnamefont {Hasegawa}}, \bibinfo {author}
  {\bibfnamefont {M.~G.}\ \bibnamefont {Silly}}, \bibinfo {author}
  {\bibfnamefont {F.}~\bibnamefont {Sirotti}}, \ and\ \bibinfo {author}
  {\bibfnamefont {I.}~\bibnamefont {Matsuda}},\ }\href {\doibase
  10.1103/PhysRevLett.108.116802} {\bibfield  {journal} {\bibinfo  {journal}
  {Physical Review Letters}\ }\textbf {\bibinfo {volume} {108}},\ \bibinfo
  {pages} {116802} (\bibinfo {year} {2012})}\BibitemShut {NoStop}%
\bibitem [{\citenamefont {Santander-Syro}\ \emph {et~al.}(2012)\citenamefont
  {Santander-Syro}, \citenamefont {Bareille}, \citenamefont {Fortuna},
  \citenamefont {Copie}, \citenamefont {Gabay}, \citenamefont {Bertran},
  \citenamefont {Taleb-Ibrahimi}, \citenamefont {{Le F{\`{e}}vre}},
  \citenamefont {Herranz}, \citenamefont {Reyren}, \citenamefont {Bibes},
  \citenamefont {Barth{\'{e}}l{\'{e}}my}, \citenamefont {Lecoeur},
  \citenamefont {Guevara},\ and\ \citenamefont
  {Rozenberg}}]{Santander-Syro2012}%
  \BibitemOpen
  \bibfield  {author} {\bibinfo {author} {\bibfnamefont {A.~F.}\ \bibnamefont
  {Santander-Syro}}, \bibinfo {author} {\bibfnamefont {C.}~\bibnamefont
  {Bareille}}, \bibinfo {author} {\bibfnamefont {F.}~\bibnamefont {Fortuna}},
  \bibinfo {author} {\bibfnamefont {O.}~\bibnamefont {Copie}}, \bibinfo
  {author} {\bibfnamefont {M.}~\bibnamefont {Gabay}}, \bibinfo {author}
  {\bibfnamefont {F.}~\bibnamefont {Bertran}}, \bibinfo {author} {\bibfnamefont
  {A.}~\bibnamefont {Taleb-Ibrahimi}}, \bibinfo {author} {\bibfnamefont
  {P.}~\bibnamefont {{Le F{\`{e}}vre}}}, \bibinfo {author} {\bibfnamefont
  {G.}~\bibnamefont {Herranz}}, \bibinfo {author} {\bibfnamefont
  {N.}~\bibnamefont {Reyren}}, \bibinfo {author} {\bibfnamefont
  {M.}~\bibnamefont {Bibes}}, \bibinfo {author} {\bibfnamefont
  {A.}~\bibnamefont {Barth{\'{e}}l{\'{e}}my}}, \bibinfo {author} {\bibfnamefont
  {P.}~\bibnamefont {Lecoeur}}, \bibinfo {author} {\bibfnamefont
  {J.}~\bibnamefont {Guevara}}, \ and\ \bibinfo {author} {\bibfnamefont
  {M.~J.}\ \bibnamefont {Rozenberg}},\ }\href {\doibase
  10.1103/PhysRevB.86.121107} {\bibfield  {journal} {\bibinfo  {journal}
  {Physical Review B - Condensed Matter and Materials Physics}\ }\textbf
  {\bibinfo {volume} {86}},\ \bibinfo {pages} {121107} (\bibinfo {year}
  {2012})}\BibitemShut {NoStop}%
\bibitem [{\citenamefont {Bareille}\ \emph {et~al.}(2014)\citenamefont
  {Bareille}, \citenamefont {Fortuna}, \citenamefont {R{\"{o}}del},
  \citenamefont {Bertran}, \citenamefont {Gabay}, \citenamefont {Cubelos},
  \citenamefont {Taleb-Ibrahimi}, \citenamefont {{Le F{\`{e}}vre}},
  \citenamefont {Bibes}, \citenamefont {Barth{\'{e}}l{\'{e}}my}, \citenamefont
  {Maroutian}, \citenamefont {Lecoeur}, \citenamefont {Rozenberg},\ and\
  \citenamefont {Santander-Syro}}]{Bareille2014}%
  \BibitemOpen
  \bibfield  {author} {\bibinfo {author} {\bibfnamefont {C.}~\bibnamefont
  {Bareille}}, \bibinfo {author} {\bibfnamefont {F.}~\bibnamefont {Fortuna}},
  \bibinfo {author} {\bibfnamefont {T.~C.}\ \bibnamefont {R{\"{o}}del}},
  \bibinfo {author} {\bibfnamefont {F.}~\bibnamefont {Bertran}}, \bibinfo
  {author} {\bibfnamefont {M.}~\bibnamefont {Gabay}}, \bibinfo {author}
  {\bibfnamefont {O.~H.}\ \bibnamefont {Cubelos}}, \bibinfo {author}
  {\bibfnamefont {A.}~\bibnamefont {Taleb-Ibrahimi}}, \bibinfo {author}
  {\bibfnamefont {P.}~\bibnamefont {{Le F{\`{e}}vre}}}, \bibinfo {author}
  {\bibfnamefont {M.}~\bibnamefont {Bibes}}, \bibinfo {author} {\bibfnamefont
  {A.}~\bibnamefont {Barth{\'{e}}l{\'{e}}my}}, \bibinfo {author} {\bibfnamefont
  {T.}~\bibnamefont {Maroutian}}, \bibinfo {author} {\bibfnamefont
  {P.}~\bibnamefont {Lecoeur}}, \bibinfo {author} {\bibfnamefont {M.~J.}\
  \bibnamefont {Rozenberg}}, \ and\ \bibinfo {author} {\bibfnamefont {a.~F.}\
  \bibnamefont {Santander-Syro}},\ }\href {\doibase 10.1038/srep03586}
  {\bibfield  {journal} {\bibinfo  {journal} {Scientific reports}\ }\textbf
  {\bibinfo {volume} {4}},\ \bibinfo {pages} {3586} (\bibinfo {year}
  {2014})}\BibitemShut {NoStop}%
\bibitem [{\citenamefont {R{\"{o}}del}\ \emph {et~al.}(2015)\citenamefont
  {R{\"{o}}del}, \citenamefont {Fortuna}, \citenamefont {Bertran},
  \citenamefont {Gabay}, \citenamefont {Rozenberg}, \citenamefont
  {Santander-Syro},\ and\ \citenamefont {{Le F{\`{e}}vre}}}]{Rodel2015}%
  \BibitemOpen
  \bibfield  {author} {\bibinfo {author} {\bibfnamefont {T.~C.}\ \bibnamefont
  {R{\"{o}}del}}, \bibinfo {author} {\bibfnamefont {F.}~\bibnamefont
  {Fortuna}}, \bibinfo {author} {\bibfnamefont {F.}~\bibnamefont {Bertran}},
  \bibinfo {author} {\bibfnamefont {M.}~\bibnamefont {Gabay}}, \bibinfo
  {author} {\bibfnamefont {M.~J.}\ \bibnamefont {Rozenberg}}, \bibinfo {author}
  {\bibfnamefont {A.~F.}\ \bibnamefont {Santander-Syro}}, \ and\ \bibinfo
  {author} {\bibfnamefont {P.}~\bibnamefont {{Le F{\`{e}}vre}}},\ }\href
  {\doibase 10.1103/PhysRevB.92.041106} {\bibfield  {journal} {\bibinfo
  {journal} {Physical Review B - Condensed Matter and Materials Physics}\
  }\textbf {\bibinfo {volume} {92}},\ \bibinfo {pages} {041106} (\bibinfo
  {year} {2015})}\BibitemShut {NoStop}%
\bibitem [{\citenamefont {Piper}\ \emph {et~al.}(2010)\citenamefont {Piper},
  \citenamefont {Preston}, \citenamefont {Fedorov}, \citenamefont {Cho},
  \citenamefont {Demasi},\ and\ \citenamefont {Smith}}]{Piper2010}%
  \BibitemOpen
  \bibfield  {author} {\bibinfo {author} {\bibfnamefont {L.~F.~J.}\
  \bibnamefont {Piper}}, \bibinfo {author} {\bibfnamefont {A.~R.~H.}\
  \bibnamefont {Preston}}, \bibinfo {author} {\bibfnamefont {A.}~\bibnamefont
  {Fedorov}}, \bibinfo {author} {\bibfnamefont {S.~W.}\ \bibnamefont {Cho}},
  \bibinfo {author} {\bibfnamefont {A.}~\bibnamefont {Demasi}}, \ and\ \bibinfo
  {author} {\bibfnamefont {K.~E.}\ \bibnamefont {Smith}},\ }\href {\doibase
  10.1103/PhysRevB.81.233305} {\bibfield  {journal} {\bibinfo  {journal}
  {Physical Review B - Condensed Matter and Materials Physics}\ }\textbf
  {\bibinfo {volume} {81}},\ \bibinfo {pages} {233305} (\bibinfo {year}
  {2010})}\BibitemShut {NoStop}%
\bibitem [{\citenamefont {Ozawa}\ \emph {et~al.}(2014)\citenamefont {Ozawa},
  \citenamefont {Emori}, \citenamefont {Yamamoto}, \citenamefont {Yukawa},
  \citenamefont {Yamamoto}, \citenamefont {Hobara}, \citenamefont {Fujikawa},
  \citenamefont {Sakama},\ and\ \citenamefont {Matsuda}}]{Ozawa2014}%
  \BibitemOpen
  \bibfield  {author} {\bibinfo {author} {\bibfnamefont {K.}~\bibnamefont
  {Ozawa}}, \bibinfo {author} {\bibfnamefont {M.}~\bibnamefont {Emori}},
  \bibinfo {author} {\bibfnamefont {S.}~\bibnamefont {Yamamoto}}, \bibinfo
  {author} {\bibfnamefont {R.}~\bibnamefont {Yukawa}}, \bibinfo {author}
  {\bibfnamefont {S.}~\bibnamefont {Yamamoto}}, \bibinfo {author}
  {\bibfnamefont {R.}~\bibnamefont {Hobara}}, \bibinfo {author} {\bibfnamefont
  {K.}~\bibnamefont {Fujikawa}}, \bibinfo {author} {\bibfnamefont
  {H.}~\bibnamefont {Sakama}}, \ and\ \bibinfo {author} {\bibfnamefont
  {I.}~\bibnamefont {Matsuda}},\ }\href@noop {} {\bibfield  {journal} {\bibinfo
   {journal} {J. Phys. Chem. Lett}\ }\textbf {\bibinfo {volume} {5}},\ \bibinfo
  {pages} {1953} (\bibinfo {year} {2014})}\BibitemShut {NoStop}%
\bibitem [{\citenamefont {Zhang}\ \emph {et~al.}(2013)\citenamefont {Zhang},
  \citenamefont {Egdell}, \citenamefont {Offi}, \citenamefont {Iacobucci},
  \citenamefont {Petaccia}, \citenamefont {Gorovikov},\ and\ \citenamefont
  {King}}]{Zhang2013}%
  \BibitemOpen
  \bibfield  {author} {\bibinfo {author} {\bibfnamefont {K.~H.~L.}\
  \bibnamefont {Zhang}}, \bibinfo {author} {\bibfnamefont {R.~G.}\ \bibnamefont
  {Egdell}}, \bibinfo {author} {\bibfnamefont {F.}~\bibnamefont {Offi}},
  \bibinfo {author} {\bibfnamefont {S.}~\bibnamefont {Iacobucci}}, \bibinfo
  {author} {\bibfnamefont {L.}~\bibnamefont {Petaccia}}, \bibinfo {author}
  {\bibfnamefont {S.}~\bibnamefont {Gorovikov}}, \ and\ \bibinfo {author}
  {\bibfnamefont {P.~D.~C.}\ \bibnamefont {King}},\ }\href {\doibase
  10.1103/PhysRevLett.110.056803} {\bibfield  {journal} {\bibinfo  {journal}
  {Physical Review Letters}\ }\textbf {\bibinfo {volume} {110}},\ \bibinfo
  {pages} {056803} (\bibinfo {year} {2013})}\BibitemShut {NoStop}%
\bibitem [{\citenamefont {Muff}\ \emph {et~al.}(2017)\citenamefont {Muff},
  \citenamefont {Fanciulli}, \citenamefont {Weber}, \citenamefont {Pilet},
  \citenamefont {Ristic}, \citenamefont {Wang}, \citenamefont {Plumb},
  \citenamefont {Radovic},\ and\ \citenamefont {Dil}}]{Muff2017}%
  \BibitemOpen
  \bibfield  {author} {\bibinfo {author} {\bibfnamefont {S.}~\bibnamefont
  {Muff}}, \bibinfo {author} {\bibfnamefont {M.}~\bibnamefont {Fanciulli}},
  \bibinfo {author} {\bibfnamefont {A.~P.}\ \bibnamefont {Weber}}, \bibinfo
  {author} {\bibfnamefont {N.}~\bibnamefont {Pilet}}, \bibinfo {author}
  {\bibfnamefont {Z.}~\bibnamefont {Ristic}}, \bibinfo {author} {\bibfnamefont
  {Z.}~\bibnamefont {Wang}}, \bibinfo {author} {\bibfnamefont {N.~C.}\
  \bibnamefont {Plumb}}, \bibinfo {author} {\bibfnamefont {M.}~\bibnamefont
  {Radovic}}, \ and\ \bibinfo {author} {\bibfnamefont {J.~H.}\ \bibnamefont
  {Dil}},\ }\href {\doibase 10.1016/j.apsusc.2017.05.229} {\  (\bibinfo {year}
  {2017}),\ 10.1016/j.apsusc.2017.05.229},\ \Eprint
  {http://arxiv.org/abs/1705.09495} {arXiv:1705.09495} \BibitemShut {NoStop}%
\bibitem [{\citenamefont {Piper}\ \emph {et~al.}(2008)\citenamefont {Piper},
  \citenamefont {Colakerol}, \citenamefont {King}, \citenamefont {Schleife},
  \citenamefont {Z{\'{u}}{\~{n}}iga-P{\'{e}}rez}, \citenamefont {Glans},
  \citenamefont {Learmonth}, \citenamefont {Federov}, \citenamefont {Veal},
  \citenamefont {Fuchs}, \citenamefont {Mu{\~{n}}oz-Sanjos{\'{e}}},
  \citenamefont {Bechstedt}, \citenamefont {McConville},\ and\ \citenamefont
  {Smith}}]{Piper2008}%
  \BibitemOpen
  \bibfield  {author} {\bibinfo {author} {\bibfnamefont {L.~F.~J.}\
  \bibnamefont {Piper}}, \bibinfo {author} {\bibfnamefont {L.}~\bibnamefont
  {Colakerol}}, \bibinfo {author} {\bibfnamefont {P.~D.~C.}\ \bibnamefont
  {King}}, \bibinfo {author} {\bibfnamefont {A.}~\bibnamefont {Schleife}},
  \bibinfo {author} {\bibfnamefont {J.}~\bibnamefont
  {Z{\'{u}}{\~{n}}iga-P{\'{e}}rez}}, \bibinfo {author} {\bibfnamefont {P.~A.}\
  \bibnamefont {Glans}}, \bibinfo {author} {\bibfnamefont {T.}~\bibnamefont
  {Learmonth}}, \bibinfo {author} {\bibfnamefont {A.}~\bibnamefont {Federov}},
  \bibinfo {author} {\bibfnamefont {T.~D.}\ \bibnamefont {Veal}}, \bibinfo
  {author} {\bibfnamefont {F.}~\bibnamefont {Fuchs}}, \bibinfo {author}
  {\bibfnamefont {V.}~\bibnamefont {Mu{\~{n}}oz-Sanjos{\'{e}}}}, \bibinfo
  {author} {\bibfnamefont {F.}~\bibnamefont {Bechstedt}}, \bibinfo {author}
  {\bibfnamefont {C.~F.}\ \bibnamefont {McConville}}, \ and\ \bibinfo {author}
  {\bibfnamefont {K.~E.}\ \bibnamefont {Smith}},\ }\href {\doibase
  10.1103/PhysRevB.78.165127} {\bibfield  {journal} {\bibinfo  {journal}
  {Physical Review B - Condensed Matter and Materials Physics}\ }\textbf
  {\bibinfo {volume} {78}},\ \bibinfo {pages} {165127} (\bibinfo {year}
  {2008})}\BibitemShut {NoStop}%
\bibitem [{\citenamefont {King}\ \emph {et~al.}(2010)\citenamefont {King},
  \citenamefont {Veal}, \citenamefont {McConville}, \citenamefont
  {Z{\'{u}}{\~{n}}iga-P{\'{e}}rez}, \citenamefont {Mu{\~{n}}oz-Sanjos{\'{e}}},
  \citenamefont {Hopkinson}, \citenamefont {Rienks}, \citenamefont {Jensen},\
  and\ \citenamefont {Hofmann}}]{King2010}%
  \BibitemOpen
  \bibfield  {author} {\bibinfo {author} {\bibfnamefont {P.~D.~C.}\
  \bibnamefont {King}}, \bibinfo {author} {\bibfnamefont {T.~D.}\ \bibnamefont
  {Veal}}, \bibinfo {author} {\bibfnamefont {C.~F.}\ \bibnamefont
  {McConville}}, \bibinfo {author} {\bibfnamefont {J.}~\bibnamefont
  {Z{\'{u}}{\~{n}}iga-P{\'{e}}rez}}, \bibinfo {author} {\bibfnamefont
  {V.}~\bibnamefont {Mu{\~{n}}oz-Sanjos{\'{e}}}}, \bibinfo {author}
  {\bibfnamefont {M.}~\bibnamefont {Hopkinson}}, \bibinfo {author}
  {\bibfnamefont {E.~D.~L.}\ \bibnamefont {Rienks}}, \bibinfo {author}
  {\bibfnamefont {M.~F.}\ \bibnamefont {Jensen}}, \ and\ \bibinfo {author}
  {\bibfnamefont {P.}~\bibnamefont {Hofmann}},\ }\href {\doibase
  10.1103/PhysRevLett.104.256803} {\bibfield  {journal} {\bibinfo  {journal}
  {Physical Review Letters}\ }\textbf {\bibinfo {volume} {104}},\ \bibinfo
  {pages} {256803} (\bibinfo {year} {2010})}\BibitemShut {NoStop}%
\bibitem [{\citenamefont {Mudd}\ \emph
  {et~al.}(2014{\natexlab{a}})\citenamefont {Mudd}, \citenamefont {Lee},
  \citenamefont {Mu{\~{n}}oz-Sanjos{\'{e}}}, \citenamefont
  {Z{\'{u}}{\~{n}}iga-P{\'{e}}rez}, \citenamefont {Hesp}, \citenamefont {Kahk},
  \citenamefont {Payne}, \citenamefont {Egdell},\ and\ \citenamefont
  {McConville}}]{Mudd2014a}%
  \BibitemOpen
  \bibfield  {author} {\bibinfo {author} {\bibfnamefont {J.~J.}\ \bibnamefont
  {Mudd}}, \bibinfo {author} {\bibfnamefont {T.-L.}\ \bibnamefont {Lee}},
  \bibinfo {author} {\bibfnamefont {V.}~\bibnamefont
  {Mu{\~{n}}oz-Sanjos{\'{e}}}}, \bibinfo {author} {\bibfnamefont
  {J.}~\bibnamefont {Z{\'{u}}{\~{n}}iga-P{\'{e}}rez}}, \bibinfo {author}
  {\bibfnamefont {D.}~\bibnamefont {Hesp}}, \bibinfo {author} {\bibfnamefont
  {J.~M.}\ \bibnamefont {Kahk}}, \bibinfo {author} {\bibfnamefont {D.~J.}\
  \bibnamefont {Payne}}, \bibinfo {author} {\bibfnamefont {R.~G.}\ \bibnamefont
  {Egdell}}, \ and\ \bibinfo {author} {\bibfnamefont {C.~F.}\ \bibnamefont
  {McConville}},\ }\href {\doibase 10.1103/PhysRevB.89.035203} {\bibfield
  {journal} {\bibinfo  {journal} {Physical Review B}\ }\textbf {\bibinfo
  {volume} {89}},\ \bibinfo {pages} {035203} (\bibinfo {year}
  {2014}{\natexlab{a}})}\BibitemShut {NoStop}%
\bibitem [{\citenamefont {Yu}\ \emph {et~al.}(2012)\citenamefont {Yu},
  \citenamefont {Mayer}, \citenamefont {Speaks}, \citenamefont {He},
  \citenamefont {Zhao}, \citenamefont {Hsu}, \citenamefont {Mao}, \citenamefont
  {Haller}, \citenamefont {Yu}, \citenamefont {Mayer}, \citenamefont {Speaks},
  \citenamefont {He}, \citenamefont {Zhao}, \citenamefont {Hsu}, \citenamefont
  {Mao}, \citenamefont {Haller},\ and\ \citenamefont {Walukiewicz}}]{Yu2014}%
  \BibitemOpen
  \bibfield  {author} {\bibinfo {author} {\bibfnamefont {K.~M.}\ \bibnamefont
  {Yu}}, \bibinfo {author} {\bibfnamefont {M.~A.}\ \bibnamefont {Mayer}},
  \bibinfo {author} {\bibfnamefont {D.~T.}\ \bibnamefont {Speaks}}, \bibinfo
  {author} {\bibfnamefont {H.}~\bibnamefont {He}}, \bibinfo {author}
  {\bibfnamefont {R.}~\bibnamefont {Zhao}}, \bibinfo {author} {\bibfnamefont
  {L.}~\bibnamefont {Hsu}}, \bibinfo {author} {\bibfnamefont {S.~S.}\
  \bibnamefont {Mao}}, \bibinfo {author} {\bibfnamefont {E.~E.}\ \bibnamefont
  {Haller}}, \bibinfo {author} {\bibfnamefont {K.~M.}\ \bibnamefont {Yu}},
  \bibinfo {author} {\bibfnamefont {M.~A.}\ \bibnamefont {Mayer}}, \bibinfo
  {author} {\bibfnamefont {D.~T.}\ \bibnamefont {Speaks}}, \bibinfo {author}
  {\bibfnamefont {H.}~\bibnamefont {He}}, \bibinfo {author} {\bibfnamefont
  {R.}~\bibnamefont {Zhao}}, \bibinfo {author} {\bibfnamefont {L.}~\bibnamefont
  {Hsu}}, \bibinfo {author} {\bibfnamefont {S.~S.}\ \bibnamefont {Mao}},
  \bibinfo {author} {\bibfnamefont {E.~E.}\ \bibnamefont {Haller}}, \ and\
  \bibinfo {author} {\bibfnamefont {W.}~\bibnamefont {Walukiewicz}},\ }\href
  {\doibase 10.1063/1.4729563} {\bibfield  {journal} {\bibinfo  {journal}
  {Journal of Applied Physics}\ }\textbf {\bibinfo {volume} {111}},\ \bibinfo
  {pages} {123505} (\bibinfo {year} {2012})}\BibitemShut {NoStop}%
\bibitem [{\citenamefont {Burbano}\ \emph {et~al.}(2011)\citenamefont
  {Burbano}, \citenamefont {Scanlon},\ and\ \citenamefont
  {Watson}}]{Burbano2011}%
  \BibitemOpen
  \bibfield  {author} {\bibinfo {author} {\bibfnamefont {M.}~\bibnamefont
  {Burbano}}, \bibinfo {author} {\bibfnamefont {D.~O.}\ \bibnamefont
  {Scanlon}}, \ and\ \bibinfo {author} {\bibfnamefont {G.~W.}\ \bibnamefont
  {Watson}},\ }\href {\doibase 10.1021/ja204639y} {\bibfield  {journal}
  {\bibinfo  {journal} {Journal of the American Chemical Society}\ }\textbf
  {\bibinfo {volume} {133}},\ \bibinfo {pages} {15065} (\bibinfo {year}
  {2011})}\BibitemShut {NoStop}%
\bibitem [{\citenamefont {Zu{\~{n}}iga-P{\'{e}}rez}\ \emph
  {et~al.}(2004)\citenamefont {Zu{\~{n}}iga-P{\'{e}}rez}, \citenamefont
  {Munuera}, \citenamefont {Ocal},\ and\ \citenamefont
  {Mu{\~{n}}oz-Sanjos{\'{e}}}}]{Zuniga-Perez2004}%
  \BibitemOpen
  \bibfield  {author} {\bibinfo {author} {\bibfnamefont {J.}~\bibnamefont
  {Zu{\~{n}}iga-P{\'{e}}rez}}, \bibinfo {author} {\bibfnamefont
  {C.}~\bibnamefont {Munuera}}, \bibinfo {author} {\bibfnamefont
  {C.}~\bibnamefont {Ocal}}, \ and\ \bibinfo {author} {\bibfnamefont
  {V.}~\bibnamefont {Mu{\~{n}}oz-Sanjos{\'{e}}}},\ }\href {\doibase
  10.1016/j.jcrysgro.2004.07.069} {\bibfield  {journal} {\bibinfo  {journal}
  {Journal of Crystal Growth}\ }\textbf {\bibinfo {volume} {271}},\ \bibinfo
  {pages} {223} (\bibinfo {year} {2004})}\BibitemShut {NoStop}%
\bibitem [{\citenamefont {Mudd}\ \emph
  {et~al.}(2014{\natexlab{b}})\citenamefont {Mudd}, \citenamefont {Lee},
  \citenamefont {Mu{\~{n}}oz-Sanjos{\'{e}}}, \citenamefont
  {Z{\'{u}}{\~{n}}iga-P{\'{e}}rez}, \citenamefont {Hesp}, \citenamefont {Kahk},
  \citenamefont {Payne}, \citenamefont {Egdell},\ and\ \citenamefont
  {McConville}}]{Mudd2014b}%
  \BibitemOpen
  \bibfield  {author} {\bibinfo {author} {\bibfnamefont {J.~J.}\ \bibnamefont
  {Mudd}}, \bibinfo {author} {\bibfnamefont {T.-L.}\ \bibnamefont {Lee}},
  \bibinfo {author} {\bibfnamefont {V.}~\bibnamefont
  {Mu{\~{n}}oz-Sanjos{\'{e}}}}, \bibinfo {author} {\bibfnamefont
  {J.}~\bibnamefont {Z{\'{u}}{\~{n}}iga-P{\'{e}}rez}}, \bibinfo {author}
  {\bibfnamefont {D.}~\bibnamefont {Hesp}}, \bibinfo {author} {\bibfnamefont
  {J.~M.}\ \bibnamefont {Kahk}}, \bibinfo {author} {\bibfnamefont {D.~J.}\
  \bibnamefont {Payne}}, \bibinfo {author} {\bibfnamefont {R.~G.}\ \bibnamefont
  {Egdell}}, \ and\ \bibinfo {author} {\bibfnamefont {C.~F.}\ \bibnamefont
  {McConville}},\ }\href {\doibase 10.1103/PhysRevB.89.035203} {\bibfield
  {journal} {\bibinfo  {journal} {Physical Review B}\ }\textbf {\bibinfo
  {volume} {89}},\ \bibinfo {pages} {035203} (\bibinfo {year}
  {2014}{\natexlab{b}})}\BibitemShut {NoStop}%
\bibitem [{\citenamefont {King}\ \emph
  {et~al.}(2009{\natexlab{a}})\citenamefont {King}, \citenamefont {Veal},
  \citenamefont {Jefferson}, \citenamefont {Z????iga-P??rez}, \citenamefont
  {Mu??oz-Sanjos??},\ and\ \citenamefont {McConville}}]{King2009c}%
  \BibitemOpen
  \bibfield  {author} {\bibinfo {author} {\bibfnamefont {P.~D.~C.}\
  \bibnamefont {King}}, \bibinfo {author} {\bibfnamefont {T.~D.}\ \bibnamefont
  {Veal}}, \bibinfo {author} {\bibfnamefont {P.~H.}\ \bibnamefont {Jefferson}},
  \bibinfo {author} {\bibfnamefont {J.}~\bibnamefont {Z????iga-P??rez}},
  \bibinfo {author} {\bibfnamefont {V.}~\bibnamefont {Mu??oz-Sanjos??}}, \ and\
  \bibinfo {author} {\bibfnamefont {C.~F.}\ \bibnamefont {McConville}},\ }\href
  {\doibase 10.1103/PhysRevB.79.035203} {\bibfield  {journal} {\bibinfo
  {journal} {Physical Review B - Condensed Matter and Materials Physics}\
  }\textbf {\bibinfo {volume} {79}},\ \bibinfo {pages} {1} (\bibinfo {year}
  {2009}{\natexlab{a}})}\BibitemShut {NoStop}%
\bibitem [{\citenamefont {Mudd}(2014)}]{Mudd2014c}%
  \BibitemOpen
  \bibfield  {author} {\bibinfo {author} {\bibfnamefont {J.~J.}\ \bibnamefont
  {Mudd}},\ }\emph {\bibinfo {title} {{Photoelectron Spectroscopy Investigation
  of CdO}}},\ \href@noop {} {Ph.D. thesis},\ \bibinfo  {school} {The University
  of Warwick} (\bibinfo {year} {2014})\BibitemShut {NoStop}%
\bibitem [{\citenamefont {Polack}\ \emph {et~al.}(2010)\citenamefont {Polack},
  \citenamefont {Silly}, \citenamefont {Chauvet}, \citenamefont {Lagarde},
  \citenamefont {Bergeard}, \citenamefont {Izquierdo}, \citenamefont {Chubar},
  \citenamefont {Krizmancic}, \citenamefont {Ribbens}, \citenamefont {Duval},
  \citenamefont {Basset}, \citenamefont {Kubsky},\ and\ \citenamefont
  {Sirotti}}]{Polack2010}%
  \BibitemOpen
  \bibfield  {author} {\bibinfo {author} {\bibfnamefont {F.}~\bibnamefont
  {Polack}}, \bibinfo {author} {\bibfnamefont {M.}~\bibnamefont {Silly}},
  \bibinfo {author} {\bibfnamefont {C.}~\bibnamefont {Chauvet}}, \bibinfo
  {author} {\bibfnamefont {B.}~\bibnamefont {Lagarde}}, \bibinfo {author}
  {\bibfnamefont {N.}~\bibnamefont {Bergeard}}, \bibinfo {author}
  {\bibfnamefont {M.}~\bibnamefont {Izquierdo}}, \bibinfo {author}
  {\bibfnamefont {O.}~\bibnamefont {Chubar}}, \bibinfo {author} {\bibfnamefont
  {D.}~\bibnamefont {Krizmancic}}, \bibinfo {author} {\bibfnamefont
  {M.}~\bibnamefont {Ribbens}}, \bibinfo {author} {\bibfnamefont {J.~P.}\
  \bibnamefont {Duval}}, \bibinfo {author} {\bibfnamefont {C.}~\bibnamefont
  {Basset}}, \bibinfo {author} {\bibfnamefont {S.}~\bibnamefont {Kubsky}}, \
  and\ \bibinfo {author} {\bibfnamefont {F.}~\bibnamefont {Sirotti}},\ }\href
  {\doibase 10.1063/1.3463169} {\bibfield  {journal} {\bibinfo  {journal} {AIP
  Conference Proceedings}\ }\textbf {\bibinfo {volume} {1234}},\ \bibinfo
  {pages} {185} (\bibinfo {year} {2010})}\BibitemShut {NoStop}%
\bibitem [{Cas(2011)}]{CasaXPS}%
  \BibitemOpen
  \href@noop {} {\enquote {\bibinfo {title} {{CasaXPS version 2.3.16}},}\ }
  (\bibinfo {year} {2011})\BibitemShut {NoStop}%
\bibitem [{\citenamefont {Yeh}(1993)}]{Yeh1993}%
  \BibitemOpen
  \bibfield  {author} {\bibinfo {author} {\bibfnamefont {J.}~\bibnamefont
  {Yeh}},\ }\href@noop {} {\emph {\bibinfo {title} {{Atomic Calculation of
  Photoionization Cross-Sections and Asymmetry Parameters}}}}\ (\bibinfo
  {publisher} {Gordon and Breach Science Publishers},\ \bibinfo {address}
  {Langhorne, PE},\ \bibinfo {year} {1993})\BibitemShut {NoStop}%
\bibitem [{\citenamefont {King}\ \emph
  {et~al.}(2009{\natexlab{b}})\citenamefont {King}, \citenamefont {Veal},
  \citenamefont {Schleife}, \citenamefont {Zuniga-Perez}, \citenamefont
  {Martel}, \citenamefont {Jefferson}, \citenamefont {Fuchs}, \citenamefont
  {Munoz-Sanjose}, \citenamefont {Bechstedt},\ and\ \citenamefont
  {McConville}}]{King2009b}%
  \BibitemOpen
  \bibfield  {author} {\bibinfo {author} {\bibfnamefont {P.~D.~C.}\
  \bibnamefont {King}}, \bibinfo {author} {\bibfnamefont {T.~D.}\ \bibnamefont
  {Veal}}, \bibinfo {author} {\bibfnamefont {A.}~\bibnamefont {Schleife}},
  \bibinfo {author} {\bibfnamefont {J.}~\bibnamefont {Zuniga-Perez}}, \bibinfo
  {author} {\bibfnamefont {B.}~\bibnamefont {Martel}}, \bibinfo {author}
  {\bibfnamefont {P.~H.}\ \bibnamefont {Jefferson}}, \bibinfo {author}
  {\bibfnamefont {F.}~\bibnamefont {Fuchs}}, \bibinfo {author} {\bibfnamefont
  {V.}~\bibnamefont {Munoz-Sanjose}}, \bibinfo {author} {\bibfnamefont
  {F.}~\bibnamefont {Bechstedt}}, \ and\ \bibinfo {author} {\bibfnamefont
  {C.~F.}\ \bibnamefont {McConville}},\ }\href {\doibase
  10.1103/PhysRevB.79.205205} {\bibfield  {journal} {\bibinfo  {journal}
  {Physical Review B - Condensed Matter and Materials Physics}\ }\textbf
  {\bibinfo {volume} {79}},\ \bibinfo {pages} {205205} (\bibinfo {year}
  {2009}{\natexlab{b}})}\BibitemShut {NoStop}%
\bibitem [{\citenamefont {Mudd}\ \emph
  {et~al.}(2014{\natexlab{c}})\citenamefont {Mudd}, \citenamefont {Lee},
  \citenamefont {Munoz-Sanjose}, \citenamefont {Zuniga-Perez}, \citenamefont
  {Payne}, \citenamefont {Egdell},\ and\ \citenamefont
  {McConville}}]{Mudd2014}%
  \BibitemOpen
  \bibfield  {author} {\bibinfo {author} {\bibfnamefont {J.~J.}\ \bibnamefont
  {Mudd}}, \bibinfo {author} {\bibfnamefont {T.~L.}\ \bibnamefont {Lee}},
  \bibinfo {author} {\bibfnamefont {V.}~\bibnamefont {Munoz-Sanjose}}, \bibinfo
  {author} {\bibfnamefont {J.}~\bibnamefont {Zuniga-Perez}}, \bibinfo {author}
  {\bibfnamefont {D.~J.}\ \bibnamefont {Payne}}, \bibinfo {author}
  {\bibfnamefont {R.~G.}\ \bibnamefont {Egdell}}, \ and\ \bibinfo {author}
  {\bibfnamefont {C.~F.}\ \bibnamefont {McConville}},\ }\href {\doibase
  10.1103/PhysRevB.89.165305} {\bibfield  {journal} {\bibinfo  {journal}
  {Physical Review B - Condensed Matter and Materials Physics}\ }\textbf
  {\bibinfo {volume} {89}},\ \bibinfo {pages} {165305} (\bibinfo {year}
  {2014}{\natexlab{c}})}\BibitemShut {NoStop}%
\bibitem [{\citenamefont {King}\ \emph {et~al.}(2008)\citenamefont {King},
  \citenamefont {Veal},\ and\ \citenamefont {McConville}}]{King2008a}%
  \BibitemOpen
  \bibfield  {author} {\bibinfo {author} {\bibfnamefont {P.~D.~C.}\
  \bibnamefont {King}}, \bibinfo {author} {\bibfnamefont {T.~D.}\ \bibnamefont
  {Veal}}, \ and\ \bibinfo {author} {\bibfnamefont {C.~F.}\ \bibnamefont
  {McConville}},\ }\href {\doibase 10.1103/PhysRevB.77.125305} {\bibfield
  {journal} {\bibinfo  {journal} {Physical Review B}\ }\textbf {\bibinfo
  {volume} {77}},\ \bibinfo {pages} {125305} (\bibinfo {year}
  {2008})}\BibitemShut {NoStop}%
\bibitem [{\citenamefont {Finkenrath}\ \emph {et~al.}(1969)\citenamefont
  {Finkenrath}, \citenamefont {Uhle},\ and\ \citenamefont
  {Waidelich}}]{Finkenrath1969}%
  \BibitemOpen
  \bibfield  {author} {\bibinfo {author} {\bibfnamefont {H.}~\bibnamefont
  {Finkenrath}}, \bibinfo {author} {\bibfnamefont {N.}~\bibnamefont {Uhle}}, \
  and\ \bibinfo {author} {\bibfnamefont {W.}~\bibnamefont {Waidelich}},\ }\href
  {\doibase 10.1016/0038-1098(69)90681-4} {\bibfield  {journal} {\bibinfo
  {journal} {Solid State Communications}\ }\textbf {\bibinfo {volume} {7}},\
  \bibinfo {pages} {11} (\bibinfo {year} {1969})}\BibitemShut {NoStop}%
\bibitem [{\citenamefont {Jefferson}\ \emph {et~al.}(2008)\citenamefont
  {Jefferson}, \citenamefont {Hatfield}, \citenamefont {Veal}, \citenamefont
  {King}, \citenamefont {McConville}, \citenamefont
  {Z{\'{u}}{\~{n}}iga-P{\'{e}}rez},\ and\ \citenamefont
  {Mu{\~{n}}oz-Sanjos{\'{e}}}}]{Jefferson2008}%
  \BibitemOpen
  \bibfield  {author} {\bibinfo {author} {\bibfnamefont {P.~H.}\ \bibnamefont
  {Jefferson}}, \bibinfo {author} {\bibfnamefont {S.~A.}\ \bibnamefont
  {Hatfield}}, \bibinfo {author} {\bibfnamefont {T.~D.}\ \bibnamefont {Veal}},
  \bibinfo {author} {\bibfnamefont {P.~D.~C.}\ \bibnamefont {King}}, \bibinfo
  {author} {\bibfnamefont {C.~F.}\ \bibnamefont {McConville}}, \bibinfo
  {author} {\bibfnamefont {J.}~\bibnamefont {Z{\'{u}}{\~{n}}iga-P{\'{e}}rez}},
  \ and\ \bibinfo {author} {\bibfnamefont {V.}~\bibnamefont
  {Mu{\~{n}}oz-Sanjos{\'{e}}}},\ }\href {\doibase 10.1063/1.2833269} {\bibfield
   {journal} {\bibinfo  {journal} {Applied Physics Letters}\ }\textbf {\bibinfo
  {volume} {92}},\ \bibinfo {pages} {022101} (\bibinfo {year}
  {2008})}\BibitemShut {NoStop}%
\bibitem [{\citenamefont {Demchenko}\ \emph {et~al.}(2010)\citenamefont
  {Demchenko}, \citenamefont {Denlinger}, \citenamefont {Chernyshova},
  \citenamefont {Yu}, \citenamefont {Speaks}, \citenamefont {Olalde-Velasco},
  \citenamefont {Hemmers}, \citenamefont {Walukiewicz}, \citenamefont
  {Derkachova},\ and\ \citenamefont {Lawniczak-Jablonska}}]{Demchenko2010}%
  \BibitemOpen
  \bibfield  {author} {\bibinfo {author} {\bibfnamefont {I.~N.}\ \bibnamefont
  {Demchenko}}, \bibinfo {author} {\bibfnamefont {J.~D.}\ \bibnamefont
  {Denlinger}}, \bibinfo {author} {\bibfnamefont {M.}~\bibnamefont
  {Chernyshova}}, \bibinfo {author} {\bibfnamefont {K.~M.}\ \bibnamefont {Yu}},
  \bibinfo {author} {\bibfnamefont {D.~T.}\ \bibnamefont {Speaks}}, \bibinfo
  {author} {\bibfnamefont {P.}~\bibnamefont {Olalde-Velasco}}, \bibinfo
  {author} {\bibfnamefont {O.}~\bibnamefont {Hemmers}}, \bibinfo {author}
  {\bibfnamefont {W.}~\bibnamefont {Walukiewicz}}, \bibinfo {author}
  {\bibfnamefont {A.}~\bibnamefont {Derkachova}}, \ and\ \bibinfo {author}
  {\bibfnamefont {K.}~\bibnamefont {Lawniczak-Jablonska}},\ }\href {\doibase
  10.1103/PhysRevB.82.075107} {\bibfield  {journal} {\bibinfo  {journal}
  {Physical Review B - Condensed Matter and Materials Physics}\ }\textbf
  {\bibinfo {volume} {82}},\ \bibinfo {pages} {1} (\bibinfo {year}
  {2010})}\BibitemShut {NoStop}%
\bibitem [{\citenamefont {Piper}\ \emph {et~al.}(2007)\citenamefont {Piper},
  \citenamefont {Jefferson}, \citenamefont {Veal}, \citenamefont {McConville},
  \citenamefont {Zu{\~{n}}iga-P{\'{e}}rez},\ and\ \citenamefont
  {Mu{\~{n}}oz-Sanjos{\'{e}}}}]{Piper2007}%
  \BibitemOpen
  \bibfield  {author} {\bibinfo {author} {\bibfnamefont {L.~F.~J.}\
  \bibnamefont {Piper}}, \bibinfo {author} {\bibfnamefont {P.~H.}\ \bibnamefont
  {Jefferson}}, \bibinfo {author} {\bibfnamefont {T.~D.}\ \bibnamefont {Veal}},
  \bibinfo {author} {\bibfnamefont {C.~F.}\ \bibnamefont {McConville}},
  \bibinfo {author} {\bibfnamefont {J.}~\bibnamefont
  {Zu{\~{n}}iga-P{\'{e}}rez}}, \ and\ \bibinfo {author} {\bibfnamefont
  {V.}~\bibnamefont {Mu{\~{n}}oz-Sanjos{\'{e}}}},\ }\href {\doibase
  10.1016/j.spmi.2007.04.029} {\bibfield  {journal} {\bibinfo  {journal}
  {Superlattices and Microstructures}\ }\textbf {\bibinfo {volume} {42}},\
  \bibinfo {pages} {197} (\bibinfo {year} {2007})}\BibitemShut {NoStop}%
\bibitem [{\citenamefont {Hammond}\ \emph {et~al.}(1975)\citenamefont
  {Hammond}, \citenamefont {Gaarenstroom},\ and\ \citenamefont
  {Winograd}}]{Hammond1975}%
  \BibitemOpen
  \bibfield  {author} {\bibinfo {author} {\bibfnamefont {J.~S.}\ \bibnamefont
  {Hammond}}, \bibinfo {author} {\bibfnamefont {S.~W.}\ \bibnamefont
  {Gaarenstroom}}, \ and\ \bibinfo {author} {\bibfnamefont {N.}~\bibnamefont
  {Winograd}},\ }\href {\doibase 10.1021/ac60363a019} {\bibfield  {journal}
  {\bibinfo  {journal} {Analytical Chemistry}\ }\textbf {\bibinfo {volume}
  {47}},\ \bibinfo {pages} {2193} (\bibinfo {year} {1975})}\BibitemShut
  {NoStop}%
\bibitem [{\citenamefont {Maticiuc}\ \emph {et~al.}(2017)\citenamefont
  {Maticiuc}, \citenamefont {Katerski}, \citenamefont {Danilson}, \citenamefont
  {Krunks},\ and\ \citenamefont {Hiie}}]{Maticiuc2017}%
  \BibitemOpen
  \bibfield  {author} {\bibinfo {author} {\bibfnamefont {N.}~\bibnamefont
  {Maticiuc}}, \bibinfo {author} {\bibfnamefont {A.}~\bibnamefont {Katerski}},
  \bibinfo {author} {\bibfnamefont {M.}~\bibnamefont {Danilson}}, \bibinfo
  {author} {\bibfnamefont {M.}~\bibnamefont {Krunks}}, \ and\ \bibinfo {author}
  {\bibfnamefont {J.}~\bibnamefont {Hiie}},\ }\href {\doibase
  10.1016/j.solmat.2016.10.040} {\bibfield  {journal} {\bibinfo  {journal}
  {Solar Energy Materials and Solar Cells}\ }\textbf {\bibinfo {volume}
  {160}},\ \bibinfo {pages} {211} (\bibinfo {year} {2017})}\BibitemShut
  {NoStop}%
\bibitem [{\citenamefont {Vogt}\ \emph {et~al.}(2015)\citenamefont {Vogt},
  \citenamefont {Gengenbach}, \citenamefont {Chang}, \citenamefont {Knowles},\
  and\ \citenamefont {Chaffee}}]{Vogt2015}%
  \BibitemOpen
  \bibfield  {author} {\bibinfo {author} {\bibfnamefont {C.}~\bibnamefont
  {Vogt}}, \bibinfo {author} {\bibfnamefont {T.}~\bibnamefont {Gengenbach}},
  \bibinfo {author} {\bibfnamefont {S.}~\bibnamefont {Chang}}, \bibinfo
  {author} {\bibfnamefont {G.}~\bibnamefont {Knowles}}, \ and\ \bibinfo
  {author} {\bibfnamefont {A.}~\bibnamefont {Chaffee}},\ }\href {\doibase
  10.1039/c4ta07085b} {\bibfield  {journal} {\bibinfo  {journal} {Journal of
  Materials Chemistry A}\ }\textbf {\bibinfo {volume} {3}},\ \bibinfo {pages}
  {5162} (\bibinfo {year} {2015})}\BibitemShut {NoStop}%
\bibitem [{\citenamefont {King}\ \emph
  {et~al.}(2009{\natexlab{c}})\citenamefont {King}, \citenamefont {Veal},
  \citenamefont {Jefferson}, \citenamefont {Zuniga-Perez}, \citenamefont
  {Munoz-Sanjose},\ and\ \citenamefont {McConville}}]{King2009a}%
  \BibitemOpen
  \bibfield  {author} {\bibinfo {author} {\bibfnamefont {P.~D.~C.}\
  \bibnamefont {King}}, \bibinfo {author} {\bibfnamefont {T.~D.}\ \bibnamefont
  {Veal}}, \bibinfo {author} {\bibfnamefont {P.~H.}\ \bibnamefont {Jefferson}},
  \bibinfo {author} {\bibfnamefont {J.}~\bibnamefont {Zuniga-Perez}}, \bibinfo
  {author} {\bibfnamefont {V.}~\bibnamefont {Munoz-Sanjose}}, \ and\ \bibinfo
  {author} {\bibfnamefont {C.~F.}\ \bibnamefont {McConville}},\ }\href
  {\doibase 10.1103/PhysRevB.79.035203} {\bibfield  {journal} {\bibinfo
  {journal} {Physical Review B - Condensed Matter and Materials Physics}\
  }\textbf {\bibinfo {volume} {79}},\ \bibinfo {pages} {035203} (\bibinfo
  {year} {2009}{\natexlab{c}})}\BibitemShut {NoStop}%
\bibitem [{\citenamefont {Berthold}\ \emph {et~al.}(2016)\citenamefont
  {Berthold}, \citenamefont {Rombach}, \citenamefont {Stauden}, \citenamefont
  {Polyakov}, \citenamefont {Cimalla}, \citenamefont {Krischok}, \citenamefont
  {Bierwagen},\ and\ \citenamefont {Himmerlich}}]{Berthold2016}%
  \BibitemOpen
  \bibfield  {author} {\bibinfo {author} {\bibfnamefont {T.}~\bibnamefont
  {Berthold}}, \bibinfo {author} {\bibfnamefont {J.}~\bibnamefont {Rombach}},
  \bibinfo {author} {\bibfnamefont {T.}~\bibnamefont {Stauden}}, \bibinfo
  {author} {\bibfnamefont {V.}~\bibnamefont {Polyakov}}, \bibinfo {author}
  {\bibfnamefont {V.}~\bibnamefont {Cimalla}}, \bibinfo {author} {\bibfnamefont
  {S.}~\bibnamefont {Krischok}}, \bibinfo {author} {\bibfnamefont
  {O.}~\bibnamefont {Bierwagen}}, \ and\ \bibinfo {author} {\bibfnamefont
  {M.}~\bibnamefont {Himmerlich}},\ }\href {\doibase 10.1063/1.4972474}
  {\bibfield  {journal} {\bibinfo  {journal} {Journal of Applied Physics}\
  }\textbf {\bibinfo {volume} {120}},\ \bibinfo {pages} {245301} (\bibinfo
  {year} {2016})}\BibitemShut {NoStop}%
\bibitem [{\citenamefont {Thomas}\ and\ \citenamefont
  {Syres}(2012)}]{Thomas2012}%
  \BibitemOpen
  \bibfield  {author} {\bibinfo {author} {\bibfnamefont {A.~G.}\ \bibnamefont
  {Thomas}}\ and\ \bibinfo {author} {\bibfnamefont {K.~L.}\ \bibnamefont
  {Syres}},\ }\href {\doibase 10.1039/c2cs35057b} {\bibfield  {journal}
  {\bibinfo  {journal} {Chemical Society Reviews}\ }\textbf {\bibinfo {volume}
  {41}},\ \bibinfo {pages} {4207} (\bibinfo {year} {2012})}\BibitemShut
  {NoStop}%
\bibitem [{\citenamefont {Ozawa}\ and\ \citenamefont {Mase}(2011)}]{Ozawa2011}%
  \BibitemOpen
  \bibfield  {author} {\bibinfo {author} {\bibfnamefont {K.}~\bibnamefont
  {Ozawa}}\ and\ \bibinfo {author} {\bibfnamefont {K.}~\bibnamefont {Mase}},\
  }\href {\doibase 10.1103/PhysRevB.83.125406} {\bibfield  {journal} {\bibinfo
  {journal} {Phys. Rev. B}\ }\textbf {\bibinfo {volume} {83}},\ \bibinfo
  {pages} {125406} (\bibinfo {year} {2011})}\BibitemShut {NoStop}%
\bibitem [{\citenamefont {McCluskey}\ and\ \citenamefont
  {Jokela}(2009)}]{McCluskey2009}%
  \BibitemOpen
  \bibfield  {author} {\bibinfo {author} {\bibfnamefont {M.~D.}\ \bibnamefont
  {McCluskey}}\ and\ \bibinfo {author} {\bibfnamefont {S.~J.}\ \bibnamefont
  {Jokela}},\ }\href {\doibase 10.1063/1.3216464} {\bibfield  {journal}
  {\bibinfo  {journal} {Journal of Applied Physics}\ }\textbf {\bibinfo
  {volume} {106}} (\bibinfo {year} {2009}),\ 10.1063/1.3216464}\BibitemShut
  {NoStop}%
\bibitem [{\citenamefont {Brillson}\ \emph {et~al.}(2007)\citenamefont
  {Brillson}, \citenamefont {Mosbacker}, \citenamefont {Hetzer}, \citenamefont
  {Strzhemechny}, \citenamefont {Jessen}, \citenamefont {Look}, \citenamefont
  {Cantwell}, \citenamefont {Zhang},\ and\ \citenamefont
  {Song}}]{Brillson2007}%
  \BibitemOpen
  \bibfield  {author} {\bibinfo {author} {\bibfnamefont {L.~J.}\ \bibnamefont
  {Brillson}}, \bibinfo {author} {\bibfnamefont {H.~L.}\ \bibnamefont
  {Mosbacker}}, \bibinfo {author} {\bibfnamefont {M.~J.}\ \bibnamefont
  {Hetzer}}, \bibinfo {author} {\bibfnamefont {Y.}~\bibnamefont
  {Strzhemechny}}, \bibinfo {author} {\bibfnamefont {G.~H.}\ \bibnamefont
  {Jessen}}, \bibinfo {author} {\bibfnamefont {D.~C.}\ \bibnamefont {Look}},
  \bibinfo {author} {\bibfnamefont {G.}~\bibnamefont {Cantwell}}, \bibinfo
  {author} {\bibfnamefont {J.}~\bibnamefont {Zhang}}, \ and\ \bibinfo {author}
  {\bibfnamefont {J.~J.}\ \bibnamefont {Song}},\ }\href {\doibase
  10.1063/1.2711536} {\bibfield  {journal} {\bibinfo  {journal} {Applied
  Physics Letters}\ }\textbf {\bibinfo {volume} {90}},\ \bibinfo {pages} {2005}
  (\bibinfo {year} {2007})}\BibitemShut {NoStop}%
\bibitem [{\citenamefont {Lord}\ \emph {et~al.}(2017)\citenamefont {Lord},
  \citenamefont {Evans}, \citenamefont {Barnett}, \citenamefont {Allen},
  \citenamefont {Barron},\ and\ \citenamefont {Wilks}}]{Lord2017}%
  \BibitemOpen
  \bibfield  {author} {\bibinfo {author} {\bibfnamefont {A.~M.}\ \bibnamefont
  {Lord}}, \bibinfo {author} {\bibfnamefont {J.~E.}\ \bibnamefont {Evans}},
  \bibinfo {author} {\bibfnamefont {C.~J.}\ \bibnamefont {Barnett}}, \bibinfo
  {author} {\bibfnamefont {M.~W.}\ \bibnamefont {Allen}}, \bibinfo {author}
  {\bibfnamefont {A.~R.}\ \bibnamefont {Barron}}, \ and\ \bibinfo {author}
  {\bibfnamefont {S.~P.}\ \bibnamefont {Wilks}},\ }\href {\doibase
  10.1088/1361-648X/aa7dc8} {\bibfield  {journal} {\bibinfo  {journal} {Journal
  of Physics Condensed Matter}\ }\textbf {\bibinfo {volume} {29}} (\bibinfo
  {year} {2017}),\ 10.1088/1361-648X/aa7dc8}\BibitemShut {NoStop}%
\bibitem [{\citenamefont {Nickel}(2006)}]{Nickel2006}%
  \BibitemOpen
  \bibfield  {author} {\bibinfo {author} {\bibfnamefont {N.~H.}\ \bibnamefont
  {Nickel}},\ }\href {\doibase 10.1103/PhysRevB.73.195204} {\bibfield
  {journal} {\bibinfo  {journal} {Physical Review B - Condensed Matter and
  Materials Physics}\ }\textbf {\bibinfo {volume} {73}},\ \bibinfo {pages}
  {195204} (\bibinfo {year} {2006})}\BibitemShut {NoStop}%
\bibitem [{\citenamefont {Wang}\ \emph {et~al.}(2005)\citenamefont {Wang},
  \citenamefont {Meyer}, \citenamefont {Yin}, \citenamefont {Kunat},
  \citenamefont {Langenberg}, \citenamefont {Traeger}, \citenamefont
  {Birkner},\ and\ \citenamefont {W{\"{o}}ll}}]{Wang2005a}%
  \BibitemOpen
  \bibfield  {author} {\bibinfo {author} {\bibfnamefont {Y.}~\bibnamefont
  {Wang}}, \bibinfo {author} {\bibfnamefont {B.}~\bibnamefont {Meyer}},
  \bibinfo {author} {\bibfnamefont {X.}~\bibnamefont {Yin}}, \bibinfo {author}
  {\bibfnamefont {M.}~\bibnamefont {Kunat}}, \bibinfo {author} {\bibfnamefont
  {D.}~\bibnamefont {Langenberg}}, \bibinfo {author} {\bibfnamefont
  {F.}~\bibnamefont {Traeger}}, \bibinfo {author} {\bibfnamefont
  {A.}~\bibnamefont {Birkner}}, \ and\ \bibinfo {author} {\bibfnamefont
  {C.}~\bibnamefont {W{\"{o}}ll}},\ }\href {\doibase
  10.1103/PhysRevLett.95.266104} {\bibfield  {journal} {\bibinfo  {journal}
  {Physical Review Letters}\ }\textbf {\bibinfo {volume} {95}},\ \bibinfo
  {pages} {1} (\bibinfo {year} {2005})}\BibitemShut {NoStop}%
\bibitem [{\citenamefont {Ozawa}\ and\ \citenamefont {Mase}(2010)}]{Ozawa2010}%
  \BibitemOpen
  \bibfield  {author} {\bibinfo {author} {\bibfnamefont {K.}~\bibnamefont
  {Ozawa}}\ and\ \bibinfo {author} {\bibfnamefont {K.}~\bibnamefont {Mase}},\
  }\href {\doibase 10.1103/PhysRevB.81.205322} {\bibfield  {journal} {\bibinfo
  {journal} {Physical Review B - Condensed Matter and Materials Physics}\
  }\textbf {\bibinfo {volume} {81}},\ \bibinfo {pages} {1} (\bibinfo {year}
  {2010})}\BibitemShut {NoStop}%
\bibitem [{\citenamefont {King}\ and\ \citenamefont {Veal}(2011)}]{King2011}%
  \BibitemOpen
  \bibfield  {author} {\bibinfo {author} {\bibfnamefont {P.~D.~C.}\
  \bibnamefont {King}}\ and\ \bibinfo {author} {\bibfnamefont {T.~D.}\
  \bibnamefont {Veal}},\ }\href {\doibase 10.1088/0953-8984/23/33/334214}
  {\bibfield  {journal} {\bibinfo  {journal} {Journal of Physics: Condensed
  Matter}\ }\textbf {\bibinfo {volume} {23}},\ \bibinfo {pages} {334214}
  (\bibinfo {year} {2011})}\BibitemShut {NoStop}%
\end{thebibliography}%

\end{document}